\documentclass[orsc,nonblindrev]{informs3} % current default for manuscript submission
\usepackage[colorlinks=true,citecolor=blue]{hyperref}
\renewcommand{\theARTICLETOP}{}

\renewcommand{\MANUSCRIPTNO}{}
\OneAndAHalfSpacedXI

\usepackage{natbib}
 \bibpunct[, ]{(}{)}{,}{a}{}{,}%
 \def\bibfont{\small}%
\TheoremsNumberedThrough     % Preferred (Theorem 1, Lemma 1, Theorem 2)
\ECRepeatTheorems

\def\TheoremsNumberedThrough{ %
\theoremstyle{TH}%

\newtheorem{assumption}{Assumption}
\theoremstyle{EX}

}

\EquationsNumberedThrough    % Default: (1), (2), ...
\MANUSCRIPTNO{}

\begin{document}
%%%%%%%%%%%%%%%%

%\setcounter{page}{1}
%\DOI{10.1287/xxxx.0000.0000}%

 \RUNAUTHOR{Gaonkar and Mele}

%\RUNTITLE{A model of inter-organizational network formation}

% \TITLE{Bundling Information Goods of Decreasing Value}
\TITLE{A model of inter-organizational network formation}

\ARTICLEAUTHORS{%
\AUTHOR{Shweta Gaonkar}
\AFF{Johns Hopkins University, Carey Business School, 100 International Dr, Baltimore, MD 21202, \EMAIL{gaonkar@jhu.edu}} %, \URL{}}
\AUTHOR{Angelo Mele}
\AFF{Johns Hopkins University, Carey Business School, 100 International Dr, Baltimore, MD 21202, \EMAIL{angelo.mele@jhu.edu}} %, \URL{}}
} % end of the block

\ABSTRACT{%
		How do inter-organizational networks emerge? Accounting for interdependence among ties while studying tie formation is one of the key challenges in this area of research. We address this challenge using an equilibrium framework where firms' decisions to form links with other firms are modeled as a strategic game. In this game, firms weigh the costs and benefits of establishing a relationship with other firms and form ties if their net payoffs are positive. We characterize the equilibrium networks as exponential random graphs (ERGM), and we estimate the firms' payoffs using a Bayesian approach. To demonstrate the usefulness of our approach, we apply the framework to a co-investment network of venture capital firms in the medical device industry. The equilibrium framework allows researchers to draw economic interpretation from parameter estimates of the ERGM Model.  We learn that firms rely on their joint partners (transitivity) and prefer to form ties with firms similar to themselves (homophily). These results hold after controlling for the interdependence among ties. Another, critical advantage of a structural approach is that it allows us to simulate the effects of economic shocks or policy counterfactuals. We test two such policy shocks, namely, firm entry and regulatory change. We show how new firms' entry or a regulatory shock of minimum capital requirements increase the co-investment network's density and clustering.
% Enter your abstract
}%

%\KEYWORDS{deterministic inventory theory; infinite linear programming duality;
%  existence of optimal policies; semi-Markov decision process; cyclic schedule}

%\HISTORY{This paper was
%first submitted on April 12, 1922 and has been with the authors for
%83 years for 65 revisions.}

\maketitle
%%%%%%%%%%%%%%%%%%%%%%%%%%%%%%%%%%%%%%%%%%%%%%%%%%%%%%%%%%%%%%%%%%%%%%

% Samples of sectioning (and labeling) in MNSC
% NOTE: (1) \section and \subsection do NOT end with a period
%       (2) \subsubsection and lower need end punctuation
%       (3) capitalization is as shown (title style).
%
%\section{Introduction.}\label{intro} %%1.
%\subsection{Duality and the Classical EOQ Problem.}\label{class-EOQ} %% 1.1.
%\subsection{Outline.}\label{outline1} %% 1.2.
%\subsubsection{Cyclic Schedules for the General Deterministic SMDP.}
%  \label{cyclic-schedules} %% 1.2.1
%\section{Problem Description.}\label{problemdescription} %% 2.

% Text of your paper here

\section{Introduction}

	 In a highly competitive environment, organizations rely on their partners to navigate markets and access resources \citep{Gulati1995b, PowellEtAl1996} crucial for their success \citep{Lavie2007}. One such organization is Medtronics that employs a plethora of deals to navigate competitive and emerging markets. For instance, in 2007, Medtronic partnered with Shandong Weigao Group Co. Ltd to open a research and development center in China to develop orthopedic technologies and devices for the local market. The then-president and CEO of Medtronics, Bill Hawkins, described the deal as follows: "\emph{Weigao has a broad orthopedic and trauma product line that complements Medtronic's offerings, but even more importantly, we feel we can generate synergies with their very strong presence and reputation in China. We view Weigao as an ideal strategic partner} ".\footnote{https://www.businesswire.com/news/home/20071217006232/en/Medtronic-Weigao-Announce-Joint-Venture-China} This quote highlights how organizations come together to share complementary assets and knowledge, gaining access to local markets and reputation through partners' network effects. Our goal is to provide a theoretical and empirical framework to analyze these strategic linking decisions among firms.

In light of the importance of inter-organizational ties \citep{AhujaEtAl2012}, organizational scholars have focused on examining how these relationships are established \citep{Ahuja2000, ahuja2009, chung2000, garcia2002, Gulati1999, KogutEtAl2007, li2002, rothaermel2008, hallen2008, reuer2013, stern2014, LomiPattison2006}. In particular, research on tie formation have indicated that links are path-dependent, and firms are more likely to seek future relationships from the pool of their current direct or indirect partners \citep{Gulati1995b}. Consequently, the process of network formation is endogenous, generating interdependence among ties \citep{AhujaEtAl2012}.\footnote{The empirical findings suggest that the prior ties \citep{Gulati1995b} are important in shaping future relationships, indicating that there is an interdependence among ties.} Organizational network scholars have attempted to control for the endogeneity and interdependence by using the perspective of social network analysis and recommend a structural approach to study tie formation \citep{BorgattiHalgin2011}.

In this study, we also adopt a structural approach to examine how firms establish relationships by developing a game-theoretical model that accounts for interdependence among links in predicting tie formation. The strategic game's equilibrium characterizes the interdependence of firms' linking decisions, thus generating the interdependence among ties noted in previous empirical studies \citep{AhujaEtAl2012, Gulati1995b}.
Furthermore, an equilibrium model provides a disciplined approach to understanding the network formation process, allowing the researchers to simulate counterfactual policy experiments that predict the changes in network structure in response to any change in the environment surrounding the firms. The ability to predict network structure's emergence allows management scholars to deduce important managerial implications based on firms' strategic decisions or changes in regulation. 

We demonstrate the structural model's use by applying this approach to a network of co-investments among venture capital firms in the medical devices industry, using data from 1940 to 2013. Two or more venture capital firms are linked if they have co-invested in the same medical devices start-up. The network exhibits a standard core-periphery structure, where a few firms have many connections among themselves (the core), while the rest of the firms have few links (or ties) to the main core (the periphery). The structural model allows us to understand what factors lead to this network structure. The estimated results strongly support the homophily argument established by the extant literature on VC syndication networks \citep{GulatiGargiulo1999, KogutEtAl2007, zhang2017}. The results suggest that firms tend to form ties with firms similar to them in age, geographic location, and managed capital.

A key advantage of the structural equilibrium approach is that it allows researchers to run policy counterfactual experiments. Our model explicitly accounts for the equilibrium network effects on individual firms' decisions to form and end ties. When there is a policy change or an exogenous shock to the economy, firms will make an optimal decision regarding their new partnerships and alliances, thus changing their network's shape. However, firms will also respond optimally to other firms' decisions due to the policy change or exogenous shock, triggering an additional secondary adjustment of their strategy following these equilibrium effects.\footnote{Our equilibrium model takes these feedback effects into account, while other partial equilibrium models are not able to incorporate the equilibrium adjustments \citep{Lucas1976}.} We focus on two types of policy changes; 1) entry of new firms in the market; 2) imposition of a minimum capital to operate. We show how these policies change the equilibrium network configuration. In general, in response to the shocks (or policy changes), we observe increased network density and clustering. The entry of new firms or minimum capital requirements affects the degree distribution: many firms that did not have any links before the entry or regulatory change will form at least one alliance, thus making the network denser and contributing to the clustering increase. The new links are not randomly allocated, as we document an average increase in homophily: thus, our model predicts that firms will tend to form more links to similar firms after these shocks or policy changes.

	\section{Formation of Inter-organizational Ties}
	The management literature is increasingly interested in understanding how network structures emerge \citep{AhujaEtAl2012, KimEtAlERGMSMJ2015}. This line of inquiry has examined tie formation as a consequence of firms attempting to fulfill their strategic or resource needs through their partners. However, the opportunity to form ties is constrained by their social opportunities \citep{EisenhardtSchooven1996}. In other words, ties established to fulfill a firm's needs are with firms within their current or past relationships. Empirical research that has examined tie formation has focused on the aspect of a firms' resource needs that will be fulfilled through their partners \citep{chung2000, Hallen2014} or their opportunities to establish relationships with other firms \citep{Brennecke2020, GulatiEtAl2012, KangZaheer2018, ZaheerSoda2009}.  Few studies have examined network formation as a strategic outcome of all firms' seeking partners. 	
	
	  Studies that have examined tie formation in an empirical setting have described it as a dyadic or triadic level process. As a consequence of such a definition, the most prominent approach used to study tie formation among firms is a binomial model such as logit or probit. More recently, academics are increasingly concerned about the interdependence among ties and the challenges this dependence poses on the estimation of tie formation \citep{AhujaEtAl2012, KimEtAlERGMSMJ2015, mindruta2016}. Few studies have raised the issue and recommend the use of a more structural approach to estimating tie formation \citep{Sorensen2007, RosenkopfPadula2008, KimEtAlERGMSMJ2015}. A promising approach uses exponential random graph models (ERGM) to re-examine questions related to tie formation \citep{LomiPattison2006, ghosh2016, Brennecke2020, BlockEtAl2019}. In an ERGM, the interdependence and correlation among ties are explicitly modeled, and estimation controls for the endogenous network structure \citep{MadhavanEtAl2004, LomiPattison2006, LomiPallotti2012, KimEtAlERGMSMJ2015, Brennecke2020}. This modeling strategy allows the researcher to account for the complexity of strategic interactions in the context of firm networks.

	 %  \color{red} NOT SURE ABOUT THIS LAST PARAGRAPH... SEEMS HARSH However, a common critique of this approach is that it is difficult to draw succinct economic interpretations using the parameter estimates, making it challenging to implement this approach for most empirical studies.\color{black} 
	
\subsection{Methods for Predicting Tie Formation}
	
Empirical works that have studied network formation have explored these relationships as dyadic or triadic level outcomes. Accordingly, these studies have used binary choice models to analyze these dyadic- or triadic-level relationships \citep{shipilov2012}, wherein firms select partners based on observable attributes, such as their resources, specialization \citep{chung2000, rothaermel2008},  trust \citep{Gulati1995b} or their network positions \citep{ahuja2009, GulatiGargiulo1999}. However, existing research on embeddedness has shown that network ties are correlated \citep{Gulati1995b}, making each link formation decision endogenous. This stream of research has focused on how the firm's position in a preexisting network structure determines the formation of inter-organizational relationships \citep{Gulati1995b, PowellEtAl1996, GulatiEtAl2012}. These preexisting ties create path dependence in establishing new relationships with other firms because repeated interaction reduces uncertainty while increasing interdependence \citep{GulatiGargiulo1999}. In other words, when firms face competition and uncertainty, they rely on familiar partners, they can trust \citep{BeckmanEtAl2004, SorensonStuart2008}. Hence, the decision to form ties is a consequence of the resources its potential partner can offer and the underlying network structure of the focal and partner firm \citep{EisenhardtSchooven1996}.

    The challenge of estimating network formation using standard binary choice models is that they fail to account for interdependence among ties. For example, when a firm forms several links, a discrete choice model like logit or probit would consider those decisions to be independent.\footnote{For more details regarding the limitation and current methodology used in management literature, please refer to \citep{KimEtAlERGMSMJ2015, mindruta2016}. } But, the formation of several links consists of interrelated decisions if one assumes the company behaves strategically, and each link is costly. As a consequence, the links should not be considered independent in statistical analysis. This view is supported by the embeddedness literature, which has established that prior ties strongly influence the choice of future partners \citep{Gulati1995a}, showing the importance of the underlying network structure as well as the observable actors' attributes in the formation of ties. Some management researchers who study firm networks have adopted new modeling and estimation techniques to overcome these challenges. One such approach involves the use of matching models \citep{mindruta2016, FoxEtAl2018}.  However, matching models are better suited for analyzing bipartite networks, where the market can be divided into two sides. When evaluating a market in which each firm can form links to any other firm, we need a different modeling strategy. Few papers have examined the network formation process using ERGM. For example, \cite{KimEtAlERGMSMJ2015} study board interlocks formation using ERGMs and estimate the model via approximate Maximum Likelihood  \citep{Snijders2002}.

\subsection{Network Formation Models}
	
	Organizational scholars have noted that the partnerships are established through a dynamic process \citep{AhujaEtAl2012, davis2016,garcia2002, koskinen_caimo_lomi_2015, GulatiEtAl2012} wherein firms can form or end a relationship based on the net benefits associated with the ties. The net benefit is determined by firms' attributes and their underlying network structure.  The literature on tie formation has empirically examined or theorized the micro-mechanisms driving relationships such as homophily \citep{TortorielloEtAl2011}, assortative matching \citep{azoulay2017}, and reciprocity \citep{CaimoLomi2015}. However, the dynamic process of relationship building is best captured using a utility-based equilibrium model \citep{BorgattiHalgin2011} that accounts for a relationship's costs and benefits. 

	More recently, organizational scholars have added to this conversation by examining how network structures emerge \citep{BorgattiHalgin2011} using tie based network formation models.  We propose an equilibrium model where tie formation emerges as mutual choices of strategic firms weighing each partnership's costs and benefits. These costs and benefits depend on firms' attributes as well as their endogenous network of links. This enables researchers to account for the tie dependence as a byproduct of the network formation game's equilibrium. The mix of utility-based and equilibrium approach produces testable implications for homophily, triadic closure, and other network features, providing researchers with new insights regarding how firms form ties. Mainly, what micromechanisms are driving firms to form or end a tie \citep{rivera2010}, furthermore, it allows scholars to formulate and test policy counterfactuals leading to a deeper understanding of how these ties are established and change with policy shocks. 
	
	Organizational scholars have explained the emergence of networks with formal utility-based models \citep{Mele2017} and stochastic actor-oriented models (SAOM)\citep{snijders2010}. A stochastic actor-oriented model examines tie formation as a consequence of actors seeking links to maximize their payoffs. This approach captures dependence among ties by allowing actors to make changes to their outgoing (directed) links based on local networks	' configuration. SAOM models prove useful while analyzing directed ties with longitudinal data. However, the ERGM model provides a global view of tie formation and examines the emergence of network structure from a relationship's perspective instead of a focal actor. Both approaches provide frameworks to estimate tie formation while accounting for interdependence among ties. However, depending on the context and level of analysis, such as actor or tie, one approach may prove better than the other.\footnote{Please check \citep{BlockEtAl2019} for a more detailed description of the differences between SAOM and ERGM, and when to choose one of these models. } 
	
	Our approach consists of modeling the dynamics of network formation, focusing on the long-run stationary states. Thus, we can compute the equilibrium probability of network configurations in the long run and estimate the model's parameters without the need for longitudinal data. This departure from the SAOM modeling strategy has been exploited in other work \citep{koskinen_caimo_lomi_2015}.
	 Additionally, our theoretical equilibrium model provides a framework for the economic interpretation of the estimated payoffs, allowing the researcher to generate counterfactual policy simulations to forecast how changes in the underlying economic environment affect the network structure in equilibrium. 
	Our model assumes that the ties are formed or deleted by the actors' mutually agreed decisions, whose strategic incentives are characterized by the game payoffs. Therefore, our micro-founded model focuses on actors and their incentives while providing a way to interpret the ties as an outcome of strategic interactions. 
	Conveniently, the long-run stationary distribution of the network corresponds to the likelihood of an exponential random graph. This result allows us to estimate the strategic model using techniques developed for ERGMs, and it mimics similar recent modeling strategies for longitudinal network analysis proposed in the literature \citep{koskinen_caimo_lomi_2015, KoskinenLomi2013}. 	Our approach also differs from the Separable Temporal ERGMs (STERGMs) because our assumptions on the meeting process among firms allow us to estimate the stationary distribution of the model without estimation of the transition probabilities \citep{KrivitskyHandcock2014}. 
	Furthermore, the networks that maximize the likelihood and are most likely to be observed in the data correspond to our model's game-theoretical equilibrium, where no pair of firms is willing to modify their linking strategy. Finally, additional sparsity can be introduced in the network model using ideas similar to \cite{Krivitsky2015} that have been shown to have a behavioral foundation in \cite{Butts2019} and can be easily incorporated in our payoff functions.

\section{Theoretical model of network formation} 
In this section we provide an abridged model description to explain the mechanics of network formation among firms.\footnote{For a detailed description of the formal model, refer to the technical appendix.}
We model an economy populated by $n$ firms. Each firm is characterized by some observable characteristics. Firm $i$ has attributes $x_i$, such as firm size, industry, and location. For ease of exposition, we will consider only one discrete attribute in this section (e.g., industry). We describe the network among firms as a $n\times n$ matrix $g$ of zeros and ones, whose entry at row $i$ and column $j$ is $g_{ij}$; if firms $i$ and $j$ have a link, then  $g_{ij}=1$; otherwise  $g_{ij}=0$.    

\begin{figure}[h]
\caption{Visualization of a time period for the network formation game}
\includegraphics[scale=0.5]{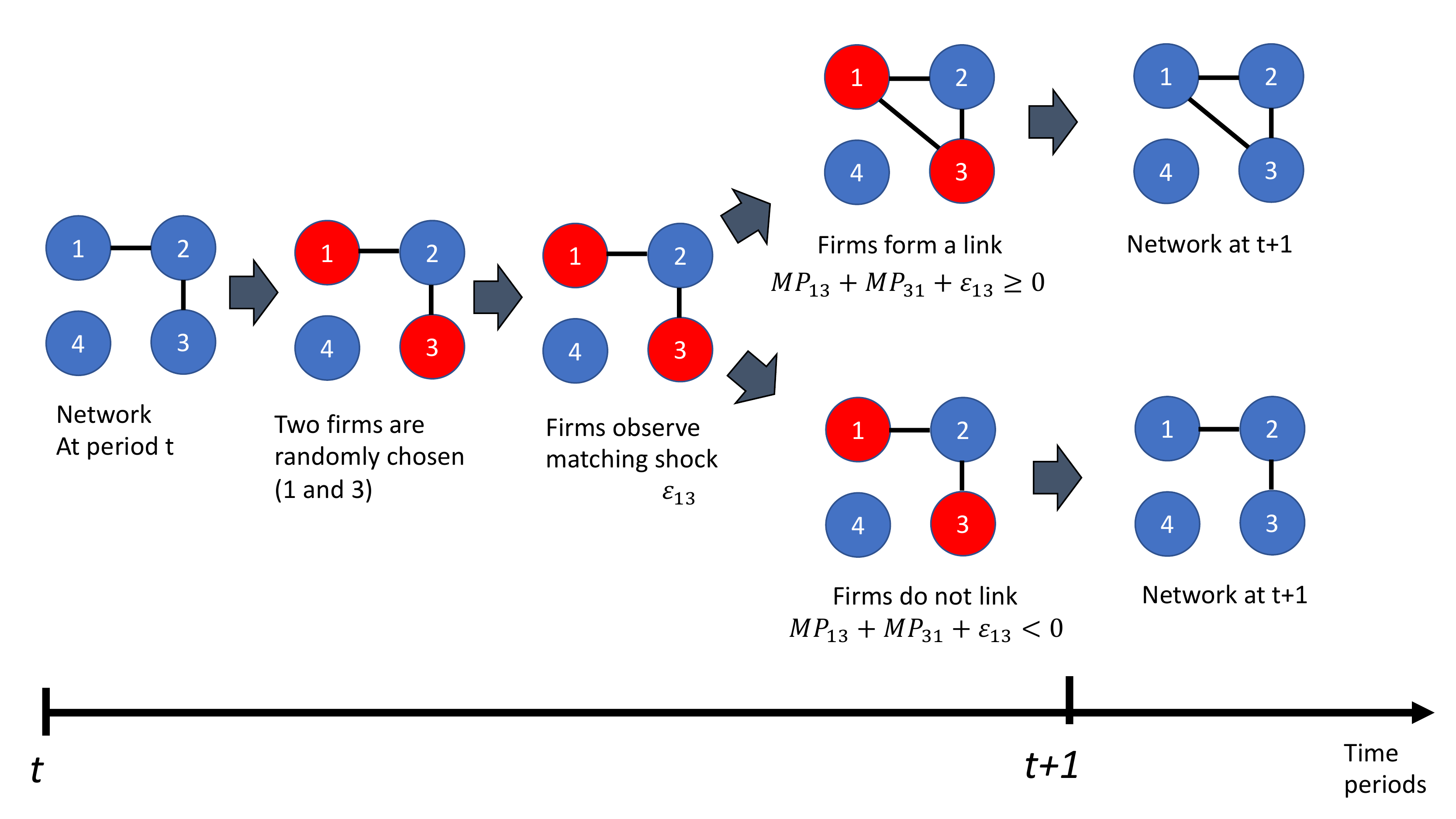}
\flushleft \scriptsize In the example we have 4 firms, represented as a blue dot. At the beginning of period $t$, the starting network is given by the left-most network, where firm 2 is linked to firm 1 and 3. The period evolves as follows: 1) two firms are randomly chosen, for example firms 1 and 3 (, highlighted as red dots)\footnote{The probability that two firms $i$ and $j$ meet and have a chance to change their link at a particular period is $\rho_{ij}>0$. This probability is \emph{positive for any pair of firms}, so that each pair of firms has an opportunity to revise their strategies. A low probability means that the pair $i$ and $j$ has rare opportunities to revise their linking strategy; on the other hand, a high probability means that $i$ and $j$ get to revise their linking strategy often over time. The probability $\rho$ is quite general, and it may depend on the structure of the network; for example, firms with many links and are popular may have a higher probability of revisions. The probability $\rho_{ij}$ can also depend on the firms' observable characteristics $x_i$ and $x_j$. For example, firms that are similar in their observable characteristics may have a higher probability of meeting. 
}; 2) firms 1 and 3 are able to observe their matching value $\epsilon_{13}$; 3) the link between 1 and 3 is formed if the surplus generated by the link is positive ($MP_{13}+MP_{31}+\varepsilon_{13}\geq 0$), otherwise the network does not change. The network is then updated and becomes the initial network at $t+1$. The game then enters next period $t+1$ and the process continues, repeating steps 1), 2) and 3). 
\label{fig:network_model_evolution} 
\end{figure}

The network is formed sequentially over time, according to the phases shown in Figure \ref{fig:network_model_evolution}. In each period of the network formation game:
\begin{enumerate}
	\item A pair of firms is randomly chosen, and they have the opportunity to update their mutual link. Updating a link involves either creation of a new tie or ending a pre-existing relationship. %Firms updating a link implies that the two firms that are not connected can choose to establish a tie. Alternatively, if the two firms with an existing link, can decide to end the relationship.
	\item Before firms decide to update their link, they observe their matching quality. The matching quality is a random variable that models unobservable characteristics affecting the outcome generated by the link. For example, it may depend on unobservable (to the researcher) long-term strategic goals of the firms, or complementarity of their research and development programs. 
	\item Firms decide jointly how to update the links, on the basis of both firms' payoffs and their matching quality. And, the economy moves to the next period. 
\end{enumerate}

The above mentioned process yields a time series of networks. Before showing how this process reaches a stationary equilibrium, we discuss the structure of the payoffs, matching quality and firms bargaining power.  

 We assume that firms evaluate different networks structures and the composition of their partners using a payoff function. The payoff has two components, deterministic and random, as in standard random utility models \citep{Heckman1978}. 
% For a firm $i$ with characteristics $x_i$ that is considering to form a link to $j$ the payoff of network $g$ is 
%\begin{equation}
%\text{Payoff of firm }i = \underset{\text{deterministic}}{\underbrace{U_i(g,x,\theta)}} + \underset{\text{random}}{\underbrace{\varepsilon_{i}}}
%\end{equation}
% In our network setting, the random component corresponds to the matching quality.
The deterministic part of payoff for firm $i$ with a network configuration $g$, firm attributes $x$, and the parameters $\theta=(\alpha_0,\alpha_1,\beta, \gamma)$ is given by the sum of net benefits of each link
%\begin{equation}
%U_{i}(g,x;\alpha,\beta,\gamma) 
%=\sum_{j=1}^n g_{ij}\left[ \underset{\text{net benefit}}{\underbrace{u(x_i,x_j;\alpha)}} + \underset{\text{popularity of } j}{\underbrace{\beta \sum_{r\neq i,j}^{n} g_{jr} }} +
%\underset{\text{common partners}}{\underbrace{\gamma \sum_{r\neq i,j}^{n} g_{jr}g_{ri} }} \right].
%\end{equation}
%\begin{equation}
%U_{i}(g,x;\alpha_0,\alpha_1,\beta,\gamma) 
%=\sum_{j=1}^n g_{ij}\left[ \underset{\text{cost}}{\underbrace{\alpha_0}} + \underset{\text{homophily/heterophily}}{\underbrace{ \alpha_1 sametype_{ij}}} + \underset{\text{popularity of } j}{\underbrace{\beta \sum_{r\neq i,j}^{n} g_{jr} }} +
%\underset{\text{common partners}}{\underbrace{\gamma \sum_{r\neq i,j}^{n} g_{jr}g_{ri} }} \right].
%\end{equation}
\begin{equation}
U_{i}(g,x;\theta) 
=\sum_{j=1}^n g_{ij}\left[ \underset{\text{cost}}{\underbrace{\alpha_0}} + \underset{\text{homophily/heterophily}}{\underbrace{ \alpha_1 \cdot sametype_{ij}}} + \underset{\text{popularity of } j}{\underbrace{\beta \cdot pop_j }} +
\underset{\text{common partners}}{\underbrace{\gamma \cdot common_{ij} }} \right].
\end{equation}
%\begin{equation}
%U_{i}(g,x;\alpha,\beta,\gamma) 
%=\alpha_0 \sum_{j=1}^n g_{ij} + \alpha_1 \sum_{j=1}^n g_{ij} sametype_{ij} + \beta \sum_{j=1}^n g_{ij} d_{j} + \gamma \sum_{j=1}^n g_{ij} \tau_{ij} .
%\end{equation}

%\begin{equation}
%U_{i}(g,x;\alpha,\beta,\gamma) = \sum_{j=1}^{n} g_{ij}MP_{ij}(g,x,\theta) 
%\end{equation}

This payoff is a weighted sum of all links formed by the ego -- firm $i$ -- where the weight is the function in the braket and corresponds to the costs and benefits generated by each link.
Each additional link has a marginal cost $\alpha_0$, and benefits accrued by homophily, popularity and shared partners. The variable $sametype_{ij}$ is an indicator equal to one if both firms -- $i$ and $j$ -- are of the same type ($x_i=x_j$), and zero otherwise. And $\alpha_1$ captures the payoff for homophily, or how much the firm values homophily. The variable $pop_{j}$ is the number of links of the alter $j$, or its popularity;  parameter $\beta$ is the value of each additional link of the alter, or the marginal benefit of the alter's popularity. The variable $common_{ij}$ is the number of shared partners for ego and alter; and the parameter $\gamma$ is the payoff for each additional common partner, or the marginal benefits of having a shared partner.

This payoff structure allows us to derive the marginal payoff for the ego firm $i$. The marginal payoff is the additional value received when $i$ forms a link to $j$, \emph{keeping the rest of the network fixed}. We denote the marginal payoff of $i$ for a link with $j$ as  $MP_{ij}$, 
\begin{equation}
MP_{ij} = \alpha_0 + \alpha_1 \cdot sametype_{ij} + \beta \cdot pop_j + 2\gamma \cdot common_{ij}
\label{eq:marginal_payoff_ij}
\end{equation} 
%To understand how the deterministic part of the payoff $U_i(g,x,\theta)$ is modeled, let's describe in detail the \emph{marginal} payoff of an additional link.
%If firm $i$ forms a link to firm $j$ the additional payoff of $i$ consists of three terms:
%\begin{eqnarray}
%MP_{ij}(g,x;\theta) &=& U_i(g_{ij}=1,g_{-ij},x,\theta) - U_i(g_{ij}=0,g_{-ij},x,\theta) \notag \\
%&=& \underset{\text{net benefit}}{\underbrace{u(x_i,x_j;\alpha)}} + \underset{\text{popularity of } j}{\underbrace{\beta \sum_{r\neq i,j}^{n} g_{jr} }} +
%\underset{\text{common partners}}{\underbrace{2\gamma \sum_{r\neq i,j}^{n} g_{jr}g_{ri} }}
%\label{eq:marginal_payoff_ij}
%\end{eqnarray}
%where $\theta=(\alpha,\beta,\gamma)$ are parameters to estimate.
	The first term of the marginal payoff -- $\alpha_0$ -- is the marginal cost of a link. The direct marginal benefit of a partner $j$ of the same type is $\alpha_1$; if it's positive ($\alpha_1>0$), then we have homophily as being of the same type will increase the marginal payoff and therefore the willingness of $i$ to form a link; viceversa if it's negative ($\alpha_1<0$) we have heterophily. Notice that in the more general version of the model we may have multiple characteristics and these can be continuous or discrete (see Appendix). \footnote{The general model that we use in the empirical analysis allows for homophily in some characteristics and heterophily in others.}

	The third term $\beta  $ is the marginal payoff from popularity of firm $j$. Each additional link that $j$ has formed increases $i$'s marginal payoff by $\beta$. If $\beta$ is positive, firm $i$ gets higher payoffs when it forms a link to a firm with many partnerships. For example, firm $j$ may have many links because of particular resources, such as unique technology or market power. This is a particular form of preferential attachment, where nodes are more likely to link other nodes that are popular \citep{AlbertBarabasi2002}. On the other hand, if the parameter $\beta$ is negative, then firm $i$ will get lower payoff from linking to a firm with many links. This may reflect a competition effect, where a firm $j$ with many links can devote fewer resources to each of its partner. Ultimately, the sign of $\beta$ is an empirical question.

The last term,  $2\gamma $, is the marginal payoff of triadic closure (transitivity). A new link to $j$ increases (or decreases) $i$'s payoff by an amount $2\gamma$, if  $j$ is a common partner. If $\gamma$ is positive, then having common partners gives higher payoffs. This term may be related to trust, cooperation, or social capital explanations, facilitating collaborations and enforcement of contracts and behaviors. It also captures the idea of repeated ties \citep{Gulati1995a}, wherein firms seek partners among their current relationships. If $\gamma$ is negative, having common partners will decrease the marginal payoff. The negative payoff is indicative of the fact that the current partners may lock a firm ability to seek partners outside their current network, leading to a negative impact of pre-existing ties \citep{AhujaEtAl2012}. As for $\beta$, the sign of $\gamma$ is an empirical matter.

The third and fourth term of the payoff correspond to the \emph{strategic} part of our model, because this part of the payoff depends on linking decisions made by other firms in the market. Indeed, the popularity of the alter is the sum of alter's links, $pop_j = \sum_{r=1}^{n} g_{jr}$; this it is a function of linking decisions $g_{jr}$ made by firm $j$. Analogously, the number of shared partners is $common_{ij}=\sum_{r=1, r\neq i,j}^{n} g_{jr}g_{ri}$ and reflects decisions about linking made by $j$ and other firms  $r$ different than $i$. Therefore these payoff terms describe how firm $i$ responds to the strategic decisions of other firms.

The random component of the payoffs corresponds to a matching quality. We assume that when two firms have the opportunity to revise their mutual link, they observe the value of the matching quality, which is a random variable, modeling the compatibility of the firms along attributes unobserved by the researcher. For example, the matching quality may depend on the long term strategic goals of the firms, their research and development objectives or the quality of their recent hires. We assume that deterministic and random components are additive, therefore the total payoff is the sum of deterministic and random component \citep{Heckman1978}.\\

We model the formation or deletion of a link as a \emph{cooperative decision, requiring both firms agreement}. Upon being selected with probability $\rho_{ij}>0$, firms $i$ and $j$ decide how to update their link $g_{ij}$. We assume that such a decision maximizes the joint surplus generated by the link, i.e., the sum of their marginal payoffs.  The marginal surplus generated by adding a link between firms $i$ and $j$ is $MP_{ij}+MP_{ji} $.

Furthermore, \emph{we allow firms to have different bargaining power to share the surplus of a link}. Usually, smaller and less established firms have less bargaining power than more established large firms in negotiations, for example. We model the differential bargaining power by allowing firms to transfer part of their payoff to the other firm, through side payments. Firms with higher bargaining power will receive a higher share of the surplus, through a transfer from the firm with lower bargaining power. This means that for a link between $i$ and $j$ the sum of payoffs is $MP_{ij} + transfer_{ij}+MP_{ji} - transfer_{ij} $, where $transfer_{ij}$ is the payoff transfer from firm $i$ to firm $j$.  Notice that the transfers cancel out since the amount $transfer_{ij}$ transferred from $i$ to $j$ is the same amount received by $j$ from $i$. Therefore the relevant quantities that determine whether two firms a form link are $MP_{ij}$ and $MP_{ji}$.

Finally, because the payoff includes an unobserved matching quality associated with creating or deleting a link, we assume that firms observe the random matching quality $\varepsilon_{ij}$ before deciding to form or delete a link, respectively. The matching quality models unobserved characteristics of the firms that the researcher cannot observe, but can affect firms' willingness to form a link—for example, compatibility of research programs among two firms, and long-term strategic agendas. We note that the match quality can be positive or negative. 

According to this framework, $i$ and $j$ will form a link if the sum of their payoffs and matching quality when they form the link is higher than the sum of payoffs and matching quality when the link is not created, that is
\begin{equation}
g_{ij}=1 \text{\ \ if \ \ } MP_{ij}+MP_{ji} + \varepsilon_{ij} \geq 0
\end{equation}
This structure is the same as most random utility models in econometrics and statistics and if the matching quality $\varepsilon_{ij}$ are independent and identically distributed according to a logistic distribution,  we can compute the probability that $i$ and $j$ form a link, conditioning on their characteristics and the shape of the network, as
\begin{equation}
P(g_{ij}=1\vert g,x,\theta) = \frac{\exp\left[MP_{ij}+MP_{ji}\right]}{1+\exp\left[MP_{ij}+MP_{ji}\right]}
\end{equation}

The structure of this game of network formation is such that only one link gets updated in each period. Furthermore, with high probability, the update will increase the surplus generated by that link. \cite{Mele2010a} shows that if we observe this process of link updates for a large number of periods, in the long-run the probability of observing a particular network configuration $g$  is\footnote{Technically, the sequence of graphs generated by the network formation game is a Markov chain \citep{LevinPeresWilmer2008, MeynTweedie2009}, converging to a unique stationary equilibrium distribution over networks, which we can characterize in closed-form.}
\begin{equation}
\pi(g,x;\theta) = \frac{\exp\left[Q(g,x;\theta)\right]}{c(\theta,x)}
\label{eq:statdist_text}
\end{equation}
where $\theta=(\alpha_0,\alpha_1,\beta,\gamma)$ is the vector of parameters to estimate and $Q(g,x;\theta)$ is the so-called \emph{potential function},
%\begin{eqnarray}
%Q(g,x;\theta) %&=& \sum_{i=1}^{n}\sum_{j=1}^{n}g_{ij}u(x_i,x_j;\alpha) + \frac{\beta}{2}\sum_{i=1}^{n}\sum_{j=1}^{n}\sum_{r\neq i,j}^{n}g_{ij}g_{jr} +
% \frac{2\gamma}{3}\sum_{i=1}^{n}\sum_{j=1}^{n}\sum_{r\neq i,j}^{n}g_{ij}g_{jr}g_{ri} \\
% &=& \sum_{i=1}^{n}\sum_{j=1}^{n}g_{ij}u(x_i,x_j;\alpha) + \beta t_{S}(g,x)+ \gamma t_{T}(g,x)
%\label{eq:potential_text}
%\end{eqnarray}
\begin{eqnarray}
Q(g,x;\theta)  &=& \alpha_0 \cdot links + \alpha_1 \cdot sametype + \beta  \cdot twostars + \gamma \cdot triangles
\label{eq:potential_text}
\end{eqnarray}

where $links$ is the total number of links in the network; $sametype$ is the number of links between firms of the same type; $twostars$ is the number of 2-stars in the network; and $triangles$ is the number of triangles.
The function $c(\theta,x)$ is a normalizing constant guaranteeing that the likelihood \eqref{eq:statdist_text} is a proper distribution.  

For a set of given values of $\alpha_0$, $\alpha_1$, $\beta$, and $\gamma$, the networks with the highest probability of being observed in the data are the ones that maximize the potential function $Q(g,x,\theta)$.\footnote{However, the networks that maximize the potential function \eqref{eq:potential_text} are not necessarily welfare-maximizing, as shown in the appendix. This means that the strategic equilibrium networks do not correspond to the optimal network architecture that maximizes the sum of the firms' payoffs in the economy. Indeed, when forming an additional link, each firm only considers the private costs and benefits of its decision, but does not take into account \emph{all} the externalities created for other firms. In practice, when a firm creates a new direct link, it is also creating an additional \emph{indirect} link for other firms, affecting their payoffs through the popularity and common partner components. In turn, this can increase the odds that those firms will form additional links in future periods. Our strategic equilibrium model captures this mechanism of cascading behavior through the potential function. However, the potential function does not include all the relevant externalities.}  These networks correspond to \emph{pairwise stable equilibrium networks with transfers}, where no pair of firms are willing to mutually form or delete a link \citep{MondererShapley1996, Mele2010a, Jackson2008, JacksonWatts2001, Mele2017}. Thus, the networks with the highest probabilities correspond to a well-defined game-theoretical equilibrium. 

We assume that the observed network from the data corresponds to a long-run equilibrium of the model to operationalize estimation. Under this assumption, we can use the likelihood function \eqref{eq:statdist_text} to estimate the model's parameters. However, this is challenging, because the normalizing constant $c(x,\theta)$ is intractable and cannot be evaluated even in small networks. The next section explains how we circumvent this problem.

Finally, we note that \emph{the model's equilibrium distribution corresponds to a particular instance of an exponential random graph (ERGM)}, with homophily, 2-stars, and triangles \citep{Snijders2002, CaimoFriel2010, KimEtAlERGMSMJ2015, Mele2010a}. While ERGMs are widely used by researchers and practitioners to estimate network formation in many applications, our model provides a game-theoretical equilibrium foundation that allows us to interpret the ERGM parameters as marginal payoffs of the firms.

\subsection{Understanding the mechanics of the model}
To understand the mechanics of the model and how the process of link updates lead to the equilibrium, we consider the model for several configurations of parameters' values. To hihglight the strategic aspect of our modeling strategy, we start from a model without strategic terms in the payoffs and use it as a baseline; we then introduce the strategic terms of the payoffs and see how this changes the possible network shapes in equilibrium.\\

\textbf{Model with no strategic terms.} To understand the role of the payoffs from popularity and common partners ($\beta$ and $\gamma$), let us first analyze the model without those effects, corresponding to a case where $\beta=0$ and $\gamma=0$. In this case, when two firms $i$ and $j$ meet, they have  a probability $p_{ij}^s = \frac{\exp(2\alpha_0 + 2\alpha_1 )}{1+\exp(2\alpha_0 + 2\alpha_1 )}$ of linking if they are of the same type ($sametype_{ij}=1$); and a probability  $p_{ij}^d = \frac{\exp(2\alpha_0)}{1+\exp(2\alpha_0 )}$ of linking if they are of different types ($sametype_{ij}=0$). We notice that in this model, the parameter $\alpha_0$ drives the density of the network. If $\alpha_0<0$ and relatively large, then the probability of a link between two firms is small, leading to a sparse network in equilibrium with few links. However, when $\alpha_0>0$, the network is relatively dense. Accordingly, if $\alpha_1>0$, the network in equilibrium will have most links among firms of the same type. But, if $\alpha_1<0$, most links are formed among firms of a different type. These are the basic elements of the model.\\

\textbf{Popularity effects}. Let's now consider the model including the effect of popularity, with $\gamma =0 $, but $\beta\neq 0$. In this case, when two firms $i$ and $j$ meet, they have  a probability $p_{ij}^s = \frac{\exp(2\alpha_0 + 2\alpha_1 + \beta (pop_j + pop_i))}{1+\exp(2\alpha_0 + 2\alpha_1 + \beta (pop_j + pop_i))}$ of linking if they are of the same type. If the two firms are of different types, then, their linking probablity is  $p_{ij}^d = \frac{\exp(2\alpha_0  + \beta  (pop_j + pop_i))}{1+\exp(2\alpha_0 +  \beta (pop_j + pop_i))}$ . From these probablities we can deduce that there are several cases that would lead to different network outcome. First, suppose that $\alpha_1>0$ and $\beta>0$. Here, firms will tend to form more links to firms of the same type and firms that are popular (have high degree), given all other things being equal. However, in some cases firms of different types may decide to form a link; as long as they are sufficiently  popular. To be concrete, consider the case $\alpha_1=\beta$; then, the probability of a link between $i$ and $j$ when they are of the same type and have both zero links ($pop_i=pop_j=0$) is $\frac{\exp(2\alpha_0 + 2\alpha_1  )}{1+\exp(2\alpha_0 + 2\alpha_1 )}$; if they are of different types but their degrees sum to 2 ($pop_i+pop_j= 2$), the probability of linking is the same as in the previous case, $ \frac{\exp(2\alpha_0  + 2\beta )}{1+\exp(2\alpha_0+ 2\beta  )}=\frac{\exp(2\alpha_0 + 2\alpha_1  )}{1+\exp(2\alpha_0 + 2\alpha_1 )}$. So in this example, firms may be willing to form links with different types of firms, as long as those firms are popular enough. Third, consider the case $\alpha_1>0$ and $\beta<0$. A similar argument applies, as two firms of different type with no links ($pop_i=pop_j=0$) may have higher probability of linking than two firms of same type with 4 links ($pop_i+pop_j= 4)$, depending on the magnitude of $\beta$. Finally, the equilibrium with $\alpha_1<0$ and $\beta<0$ consists of a relatively sparse network where most firms will link to firms of different type, and have few links, generating a relatively sparse network with heterophily.\\

\textbf{Transitivity effects}. Let's consider the model with $\beta=0$ and $\gamma\neq 0$. In this case, we only have transitivity effects and no popularity effects. When two firms $i$ and $j$ meet, they have  a probability $p_{ij}^s = \frac{\exp(2\alpha_0 + 2\alpha_1 + 4\gamma common_{ij})}{1+\exp(2\alpha_0 + 2\alpha_1 + 4\gamma common_{ij})}$ of linking if they are of the same type; they have a probability  $p_{ij}^d = \frac{\exp(2\alpha_0  + 4\gamma common_{ij})}{1+\exp(2\alpha_0 +  4\gamma common_{ij})}$  of linking if of different type. If $\alpha_1>0$ and $\gamma>0$, then most links will occur among firms with common partners and the same type. However, if $i$ and $j$ are of different type but have many common partners, it may still be profitable to form a link, since common partners' payoff may be larger than zero. For instance, if $\alpha_1=\gamma>0$ two firms of different types with one friend in common will have a higher probability of linking than two firms of the same type without any common partner. The case of $\alpha_1>0$ and $\gamma<0$ implies that having common partners decreases payoffs, thus the probability of linking. The equilibrium networks will be relatively sparse, as most links will occur among firms with no common partners, and the network will also have low clustering. More cases can be analyzed using the same reasoning.\\

This short discussion shows that even a simplified version of our model is rich enough to provide quite an array of possible network configurations with varying degrees of density, homophily, clustering, and the economy's firms' popularity. In the empirical section, we use a much richer specification of the payoffs to capture more realistic linking behavior and match the observed network properties. 

\subsection{The importance of the strategic equilibrium for policy analysis}

The distinction between the homophily part of the payoff -- $\alpha_0 + \alpha_1 \cdot sametype_{ij}$ -- and the strategic payoffs  -- $\beta$ and $\gamma$ -- has important \emph{policy-relevant implications}. Indeed, many observed networks display clustering, where nodes organize in more densely connected subnetworks while maintaining fewer connections across clusters \citep{Graham2014, AhujaEtAl2012}. There are at least two possible explanations for high clustering levels. First, the clusters may be the consequence of homophily, a preference to form links to similar nodes. Second, the clusters may arise because of a positive payoff from transitivity or popularity. If most of the clustering is due to payoffs from transitivity or popularity, then an economic shock that affects one of the nodes will also impact other nodes through equilibrium adjustment. However, if the clustering is mostly due to homophily, a shock that affects a node will not spread to other nodes.

The adjustment of equilibrium to potential policy shocks is handy to understand the ``what if'' questions, as in our policy counterfactuals. For example, suppose that a firm receives a negative economic shock. As a result, the firm may need to end some collaborations resulting in a change in the network's shape. If the payoffs do not include the strategic equilibrium terms for popularity and common partnerships ($\beta=0,\gamma=0$), the effect of the economic shock stops here. No other firms will update their linking strategy. 
On the other hand, if firms' payoffs are affected by the strategic terms ($\beta\neq 0,\gamma \neq 0$), then the network will change and adjust towards a new equilibrium. Going back to the policy shock (mentioned above), where firms decide to sever some links, the firms' partners will suffer a change to their payoff due to the changes in popularity and shared partners. This change means that once firms have elected to revise their strategy, they will update some of their links, either deleting or forming new ones. These updates to the links will result in other firms experiencing a change in their popularity or common partners, creating an incentive to modify their linking strategy in future periods. This chain of adjustments of firms strategies will converge to a new long-run equilibrium network, where with a high probability no firms are willing to change their links.

A model that does not include strategic or equilibrium considerations will not display these feedback effects and, therefore, would predict small effects of policy changes.\footnote{This challenge is similar to the \cite{Lucas1976} critique of structural equation models that do not incorporate equilibrium considerations when estimating the parameters.} In the terminology used by \cite{HeckmanVytlacil2007HB} our strategic equilibrium model describes the payoff parameters $(\alpha_0,\alpha_1, \beta, \gamma)$, that are \emph{policy invariant}. The idea of policy invariance implies that the payoff parameters do not change when economic conditions change by the effect of a policy or shock to the economy. Furthermore, our model includes \emph{endogenous} quantities -- such as the number of transitive triples -- that will change after an economic shock because of the equilibrium adjustments.

The model and our approach are useful in two ways. First, it allows us to \emph{test} whether the data support the strategic considerations we postulate. Indeed, after estimating the model, we can test whether the parameters are equal to zero, that is $\beta=0$ or $\gamma=0$ or both, according to standard statistical hypothesis testing. This test will inform the researcher whether the strategic considerations are essential or not. Second, our model allows us to predict how the network will reach an equilibrium. Therefore it allows us to study how the economy transitions from equilibrium to another after a policy shock.

\section{Estimation, Data and Model Specification}

\subsection{Application: Venture Capital Syndication}
We apply this methodology in the context of venture capital investments. \cite{wilson1968} defines a syndicate as a “group of individual decision-makers who must make a common decision under uncertainty, and who, as a result, will receive a joint payoff to be shared among them.” Venture capitalists operate in highly uncertain environments and tend to work alongside one another \citep{gu2014, lerner1994}. These venture syndicates allow venture capitalists to deal with risks in an uncertain environment \citep{manigart2006}. 

	Syndication networks provide access to future investment opportunities \citep{hochberg2007}, thus shaping future co-investment ties \citep{zhang2017}. As a result, venture capitalists seek new partners based on firm attributes \citep{SorensonStuart2008}, which is guided by their prior investments.   There is an interdependence among links in this context\citep{zhang2017}, creating an estimation challenge that can be dealt with using our structural model and estimation method. Hence, venture capital syndication is an apt empirical setting to use our model to examine network formation.

    We built a dataset of all venture capital firms with headquarters in the United States that invested in the medical device industry between 1940 and 2013. Data on venture capitalists come from the Venture Xpert database maintained by Thompson One. Our sample includes all the firms in this database that declare their investment type as "venture capital." This sample gives us a total of 833 firms. In our data, we have information, for each firm, about age, total capital under management, firm type, fund size, and address of headquarters.

\begin{figure}[t]
\caption{The network of venture capital co-investments in the medical device industry (1940-2013)}
\centering
\includegraphics[scale=.4]{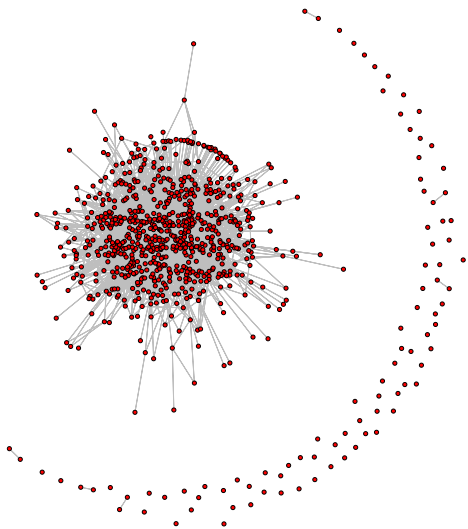}
\flushleft \scriptsize The network consists of 833 venture capital firms that have invested at least once in a company in the medical device industry between 1940 and 2013. There are 6997 co-investment links.
\label{fig:network}
\end{figure}

The network we examine in our empirical analysis is the co-investment network. If two venture capital firms have invested together in a medical device firm, then two firms are linked or have a tie. The resulting network is undirected, containing 833 nodes and 6997 links.

Figure \ref{fig:network} shows the network from the venture capitalists' data. We observe a typical core-periphery structure, where some firms are extremely connected (the core), and some firms only form a few links to the densely connected core (the periphery). We notice that some firms are isolated, as they did not form syndicates with other firms during the period considered.

\subsection{Model specification}
%We follow \cite{Mele2010a} and specify a parsimoniuous model for the payoffs. The payoff from direct links has functional form
%\begin{equation}
%u(x_i,x_j;\alpha) = \alpha_{0} + \sum_{p=1}^{P}\alpha_p f_p (x_i,x_j)
%\label{eq:direct_payoff_spec}
%\end{equation}

We have tried several specifications of the marginal payoff function, and we will focus on the following in the empirical results
\begin{eqnarray}
MP_{ij} &=& \alpha_0 + \alpha_1 sametype_{ij} + \alpha_2  \vert capital_i- capital_j\vert  \label{eq:utility_specification_empirical}\\
& + & \alpha_3  \vert age_i - age_j \vert  + \alpha_4 samestate_{ij} \notag \\
&+& \beta \cdot pop_j + 2\gamma \cdot common_{ij} \notag
\end{eqnarray}

The variable $sametype_{ij}=1$ if the venture capital firms $ij$ belongs to the same type, and $sametype_{ij}=0$ otherwise. Firms belonging to the following categories: $\lbrace$Angel Group, Bank Affiliated, Corporate PE/Venture, Endowment/Foundation or Pension Fund, Government Affiliated Program, Incubator/Development Program, Individuals, Insurance Firm Affiliate, Investment Management Firm, Private Equity Advisor or Fund of Funds, Private Equity Firm, SBIC, Service Provider, University Program$\rbrace$. The variable $capital_i$ is the (log of) capital under management at venture capital firm $i$; $age_i$ is the firm's age in 2013; and $samestate_{ij}=1$ if firms $i$ and $j$ belong to the same state, and $samestate_{ij}=0$ otherwise. The terms $pop_j$ and $common_{ij}$ are the number of links of firm $j$ (or $j$'s popularity) and the number of common partners of $i$ and $j$, respectively.

We interpret the first term $\alpha_{0}$ of payoff \eqref{eq:utility_specification_empirical} as firm $i$ cost  of forming a link. 
The remaining terms of \eqref{eq:utility_specification_empirical} represent firm's $i$ direct benefit of forming a link 
with firm $j$. \footnote{Our implicit assumption is that the cost of forming links is the same across all the firms, but their benefits may vary according to their characteristics. As explained in \cite{Mele2017} we can only identify the \emph{net benefit} of a direct connection. An alternative specification could assume that the benefit of forming a link is constant and that the cost varies with observable characteristics.}
The sign of the coefficients $(\alpha_1, \alpha_2, \alpha_3, \alpha_4, \beta, \gamma)$ is left unconstrained. The interpretation is that if we find $\alpha_1>0$, then we have homophily in types, and firms of the same type will be more likely to form links, other things being equal. Vice versa, if we find a $\alpha_1<0$, we have heterophily in firm types; thus, firms of the same type will have a lower probability of linking. The same reasoning applies to $\alpha_4$. However, for $\alpha_2$ and $\alpha_3$, the reasoning is reversed. If $\alpha_2>0$, we have heterophily in the capital, and firms with similar capital levels will have a lower probability of forming a syndicate than firms with large differences in managed capital. On the other hand, when $\alpha_2<0$, firms tend to seek partners with similar capital (homophily).

The model parameters to estimate are therefore $\theta = (\alpha_0,\alpha_1, \alpha_2, \alpha_3, \alpha_4, \beta, \gamma)$, and we denote the parameter space as $\Theta$, that is $\theta\in \Theta$.

\subsection{Estimation}
We adopt a Bayesian approach and estimate the posterior distribution of parameters, using the simulation methods developed in \citep{Mele2010a} and \citep{CaimoFriel2010}. All Bayesian analysis starts from a prior distribution $p(\theta)$ that summarizes the researcher's prior knowledge regarding the parameters of the model. We then use Bayes rule to compute the posterior distribution of the payoff parameters, $p(\theta \vert g,x )$, defined as the conditional probability distribution of the parameters $\theta = (\alpha,\beta,\gamma)$, given the data, the likelihood of the model and the prior distribution
\begin{equation}
p(\theta \vert g,x ) = \frac{\pi(g,x;\theta)p(\theta)}{\int_{\Theta} \pi(g,x;\eta)p(\eta)d\eta}
\label{eq:posterior}
\end{equation}
where $\pi(g,x;\theta)$ is the likelihood distribution \eqref{eq:statdist_text} and $\Theta$ is the set of all parameter values.

Intuitively, the posterior distribution answers the question: what parameter vectors are more likely to generate the network observed in the data, according to our strategic model, and given our prior knowledge about the parameters of the phenomenon under study? The answer to this question is a distribution of parameters that assigns a higher probability for the parameter vectors that are most likely to generate the data.

The Bayesian approach is well suited to estimating this class of models for several reasons. First, the literature on network estimation has established that some exponential random graphs models suffer from \emph{degeneracy}. Degeneracy implies that the model puts the highest probability on a small network set, usually the empty or full network. Degeneracy is problematic because, empirically, it is very unusual to observe empty or complete networks. Researchers have resolved this issue by providing alternative specifications of the model \citep{Snijders2002, RobinsEtAl2007}, by including or excluding some terms (like triangles), and by "curving" the exponential distribution \citep{Geyer1992}.

	The Bayesian approach allows the researcher to specify priors that consider how some parameter vectors imply degenerate models. A simple solution is to impose null prior probability for regions of the parameter space that generate almost empty or almost complete networks. A strength of our estimation strategy is that it does not rely on this ultimate prior choice. Indeed, the estimation algorithm that we use in this paper, called the \emph{exchange} algorithm \citep{MurrayEtAl2006, Mele2010a, CaimoFriel2010}, simulates from the posterior distribution of the parameters, giving more weight to those parameters that generate networks similar to the observed data. Therefore, the algorithm estimates a posterior where parameters that lead to degeneracy have a small or null probability. This feature of the algorithm is a remarkable property that mitigates the issue of degeneracy explained above.

Second, we bypass the problem of computing the normalizing constant $c(\theta,x)$ of the likelihood, because our algorithm recovers the posterior distribution of the parameters without evaluating the likelihood of the model. While recent advances in computational methods have allowed researchers to perform estimation in large networks \citep{ByshkinEtAl2018, StivalaEtAl2016, MeleZhu2015}, the computation of the normalizing constant is still the main challenge in estimating this class of models. The ERGM literature has developed alternative estimators, like the Markov Chain Monte Carlo Maximum Likelihood Estimator (MCMC-MLE) or the Maximum Pseudo-Likelihood Estimator (MPLE). The MCMC-MLE relies on a Monte Carlo approximation of the likelihood, through a long simulation of the model \citep{Snijders2002, GeyerThompson1992}, which estimates the parameters by maximizing the approximated likelihood. As the number of simulations grows large, the parameters estimated with this method converge to the maximum likelihood estimate \citep{GeyerThompson1992}. However, practitioners have demonstrated several difficulties with these approximations,  which may be inadequate if the simulation's chosen starting value lies too far from the (exact) maximum likelihood estimate. Besides, there are cases in which the maximum likelihood estimate does not exist; when this is the case, the simulation output is unstable, leading to unreliable inference \citep{GeyerThompson1992, Snijders2002, Butts2009}.

On the other hand, the Maximum Pseudo-Likelihood Estimator simplifies the estimation by considering the linking choices as independent \citep{Besag1974, WassermanPattison1996}. This simplification corresponds to a logit estimation with endogenous regressors. One can show that under some assumptions and as the network size becomes large, such estimates are consistent \citep{BoucherMourifie2017, Besag1974}. However, in many practical applications, the estimates are flawed and imprecise \citep{CaimoFriel2010, Mele2010a, Snijders2002}. Furthermore, standard errors are underestimated and need adjustment. \\

%\subsection{Estimation algorithm}
To estimate the model, we assume that the network we observe in the data is a stationary network drawn from the likelihood of the theoretical model (\ref{eq:statdist_text}). In many applications, the posterior distribution is available in closed-form; for example, it could be a normal distribution. However, in our case the posterior (\ref{eq:posterior}) is a \emph{doubly intractable distribution} because it contains two intractable normalizing constants. Indeed, we can rewrite the posterior as 
\begin{equation}
p(\theta \vert g,x ) = \frac{\frac{\exp\left[Q(g,x;\theta)\right]}{c(\theta,x)}p(\theta)}{\int_{\Theta} \frac{\exp\left[Q(g,x;\eta)\right]}{c(\eta,x)}p(\eta)d\eta} 
= \frac{\frac{\exp\left[Q(g,x;\theta)\right]}{c(\theta,x)}p(\theta)}{p(g,x)}
\label{eq:posterior_ergm}
\end{equation}
The first intractable constant is $p(g,x)=\int_{\Theta} \frac{\exp\left[Q(g,x;\eta)\right]}{c(\eta,x)}p(\eta)d\eta$, which corresponds to the normalizing constant of the posterior distribution. This is also called \emph{marginal likelihood} or \emph{model evidence} in the Bayesian literature \citep{GelmanEtAl2003}. The marginal likelihood is the probability of observing the network $g$ given our specific model (that is, our model's ability to explain the network in the data). This quantity is intractable because it involves a high-dimensional integration over the parameter space $\Theta$. The computation of this constant is circumvented using Metropolis-Hastings Markov Chain Monte Carlo (MCMC) algorithms \citep{LiangLiuCarrollBook2010}, simulating parameters from the posterior distribution in an iterative way.
However, in this model, we have an additional intractable normalizing constant in the likelihood
\begin{equation}
c(\theta,x)=\sum_{\omega\in\mathcal{G}}\exp\left[Q(\omega, x;\theta) \right].
\end{equation}

This is a sum over all possible networks with $n$ firms -- the set of  $\mathcal{G}$ -- containing a total of $2^{n(n-1)/2}$ terms. Even a small network with $n=20$ firms would imply a sum over  $2^{90}\approx 10^{27}$ terms, each term involving the computation of the potential function $Q(\omega,x;\theta)$. This is infeasible for most network sizes, implying that the likelihood and the posterior cannot be computed directly.\footnote{With some algebra, we can show that first and second derivatives of the likelihood also depend on the normalizing constant $c(\theta,x)$, thus making maximum likelihood estimation quite challenging.  This estimation challenge is an additional reason to prefer the Bayesian approach to this class of models' frequentist approach. However, recent computational approaches and approximations allow researchers to scale estimation to much larger networks \citep{ByshkinEtAl2018, StivalaEtAl2016, MeleZhu2015}. }

We use an \emph{approximate algorithm} that circumvents the need to compute both constants $p(g,x)$ and $c(\theta,x)$. \cite{MurrayEtAl2006} developed the original algorithm, later adapted to networks in \cite{CaimoFriel2010} and \cite{Mele2010a}. The methods is implemented in the open source package \texttt{Bergm} in the statistical software \texttt{R}. Our estimation code is available in Appendix.\footnote{Our data are proprietary, therefore we are not able to share them. The code we provide uses data generated by simulation of the model. This documentation should provide enough guidance to an empirical researcher that wants to implement this method on different data-sets.} We provide a brief description of the algorithm and its useful properties. The readers interested in technical details can find a formal description in our appendix \footnote{And, the proofs of convergence in \citep{Mele2010a}}.

We start the simulation from parameter vector $\theta^{(0)}$ and network observed in the data $g$. The simulations proceed according to the following steps:
\begin{enumerate}
	\item[STEP 1] propose a new vector of parameters $\theta^{\prime}$;
	\item[STEP 2] given the proposed parameters $\theta^{\prime}$, simulate the network formation process using the model for $R$ steps; collect the network from last step $g^{\prime}$;
	\item[STEP 3] check if the simulated network $g^{\prime}$ is similar to the network in the data $g$;
	\begin{itemize}
	\item if simulated and observed network are similar enough, accept the proposed parameter with high probability; 
	\item otherwise, reject the parameter;
	\end{itemize}
	\item[STEP 4] Repeat steps 1 to 3 for $S$ times.
\end{enumerate}

This algorithm samples parameter vectors that are most likely to generate the data. In STEP 1, we propose a parameter vector. Then, we simulate $R$ networks from the model in STEP 2. We need to choose $R$ large enough to be sure that the last simulated network $g^{\prime}$ is approximately drawn from the distribution (\ref{eq:statdist_text});\footnote{A simple rule of thumb is to use $R=n^2 \ln(n)$ where $n$ is the number of nodes of the network. However, we suggest to try different values of $R$ and show an implementation in our sample code.} we then compare this \emph{simulated} network $g^{\prime}$ with the observed network $g$, using a likelihood ratio.\footnote{\cite{Mele2010a} (Appendix B) shows that using the likelihood ratio eliminates the constant $c(\theta,x)$ and speeds up computations, thus avoiding the main computational bottleneck.} If the simulated and observed networks are similar, that is, their likelihood ratio is close to 1, then the parameter vector we have proposed is very likely to generate the network in our data and should receive high probability in the posterior. That implies that we should accept this proposed parameter with high probability. Vice versa, if the simulated and observed networks look very different and the likelihood ratio is very different from 1, the parameter we have proposed is unlikely to generate our data and thus should receive low probability in the posterior. Accordingly, we would reject this parameter vector with a high probability.

This simulation will converge to the posterior distribution of the parameters, provided that both the number of network simulations $R$, and parameter simulations $S$ are extensive. In setting the number of simulations, we follow suggestions from the applied probability literature that has established convergence speed for this class of algorithms \citep{BhamidiEtAl2011, CaimoFriel2010, Mele2010a}.

Our algorithm has several useful properties. First, the simulation strategy attenuates degeneracy problems because parameters that generate almost-empty or almost-complete networks will receive very low or null probability in the estimated posterior. Such parameters will not generate simulated networks that resemble our network data. Thus, the algorithm will reject them in favor of parameter vectors that generate networks similar to the observed data. 
Second, our estimation procedure does not require a careful choice of starting value, unlike in the MCMC-MLE method. Indeed, the exchange algorithm converges to the correct posterior distribution independently of the starting value.\footnote{ For a formal proof of convergence, see Appendix B of \citep{Mele2010a}.} By contrast, the MCMC-MLE is very sensitive to the initial parameter value, sometimes providing very poor (local) approximations to the likelihood.

Third, the software for estimation is open source and available in the statistical package R. We use the package \texttt{Bergm}, developed by \cite{CaimoFriel2010}, to perform our analysis. We also provide a sample code with guidance in the Appendix. Finally, we believe that the combination of recent advances in computation \citep{ByshkinEtAl2018, StivalaEtAl2016, MeleZhu2015} and Bayesian methods is a promising avenue for estimation of this class of models.

\section{Empirical Results}

\subsection{Descriptive statistics}
\begin{figure}[!h]
\caption{Distribution of location and types of firms in the sample}
\includegraphics[scale = 0.27]{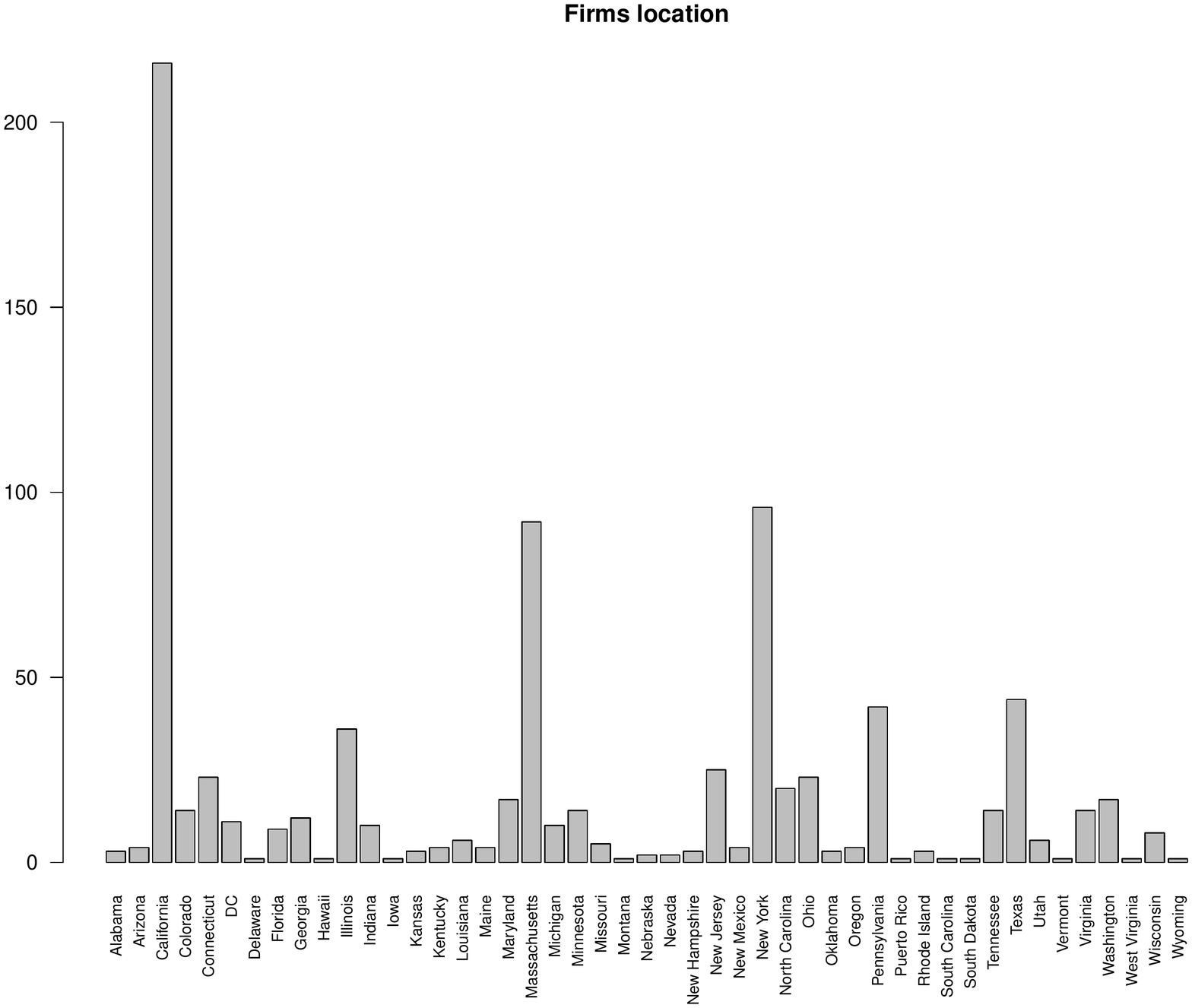}
\includegraphics[scale = 0.3]{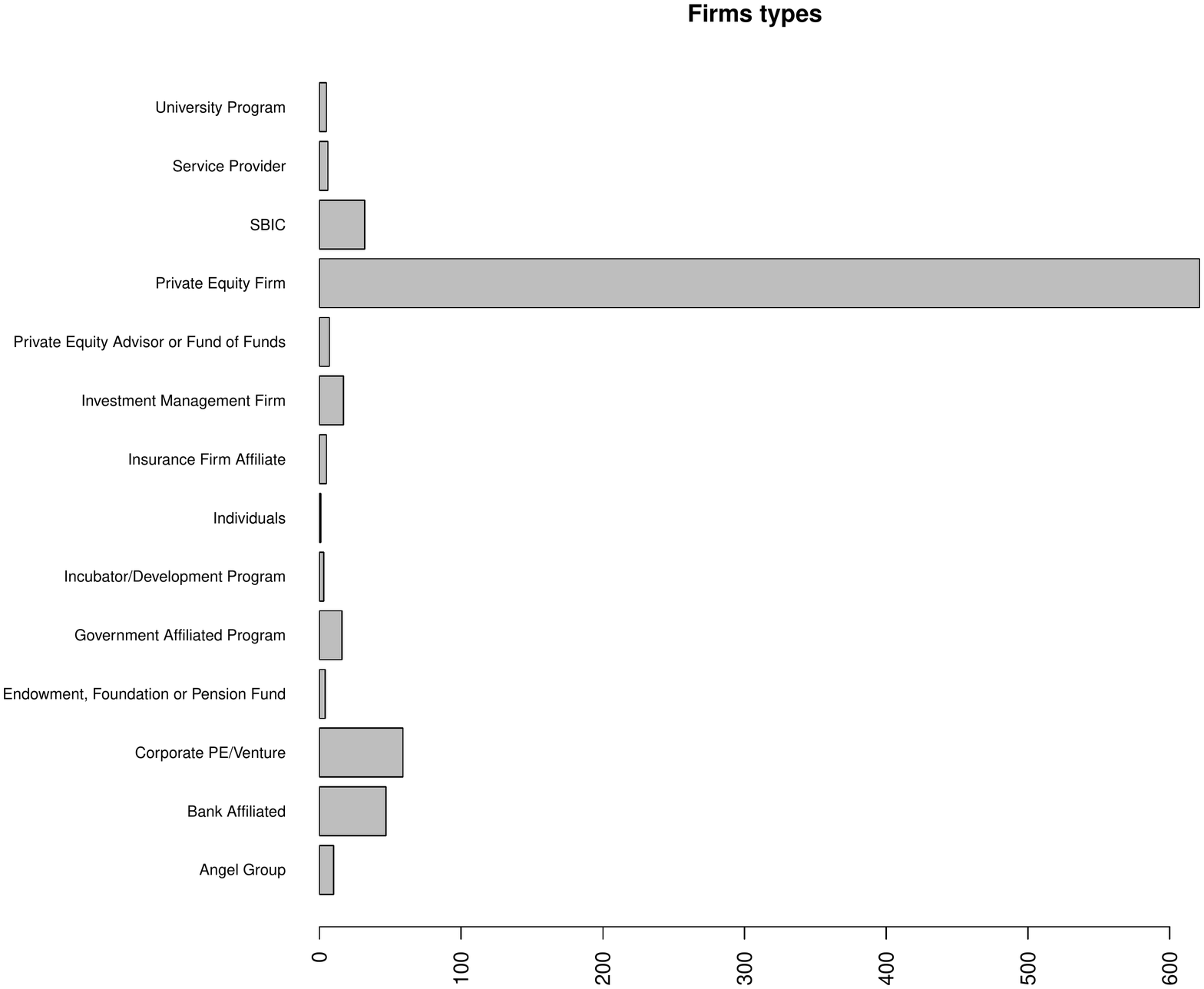}
\label{fig:location_plot}
\end{figure}

Our network contains 833 venture capital firms and 6997 co-investment partnerships. As shown in Figure \ref{fig:location_plot}, most of our firms are located in California, Massachussets, and New York (left). Most companies in our data are private equity firms (right).

\begin{figure}
\caption{Distribution of capital managed (in logs) and age in the sample}
\centering
\includegraphics[scale=.27]{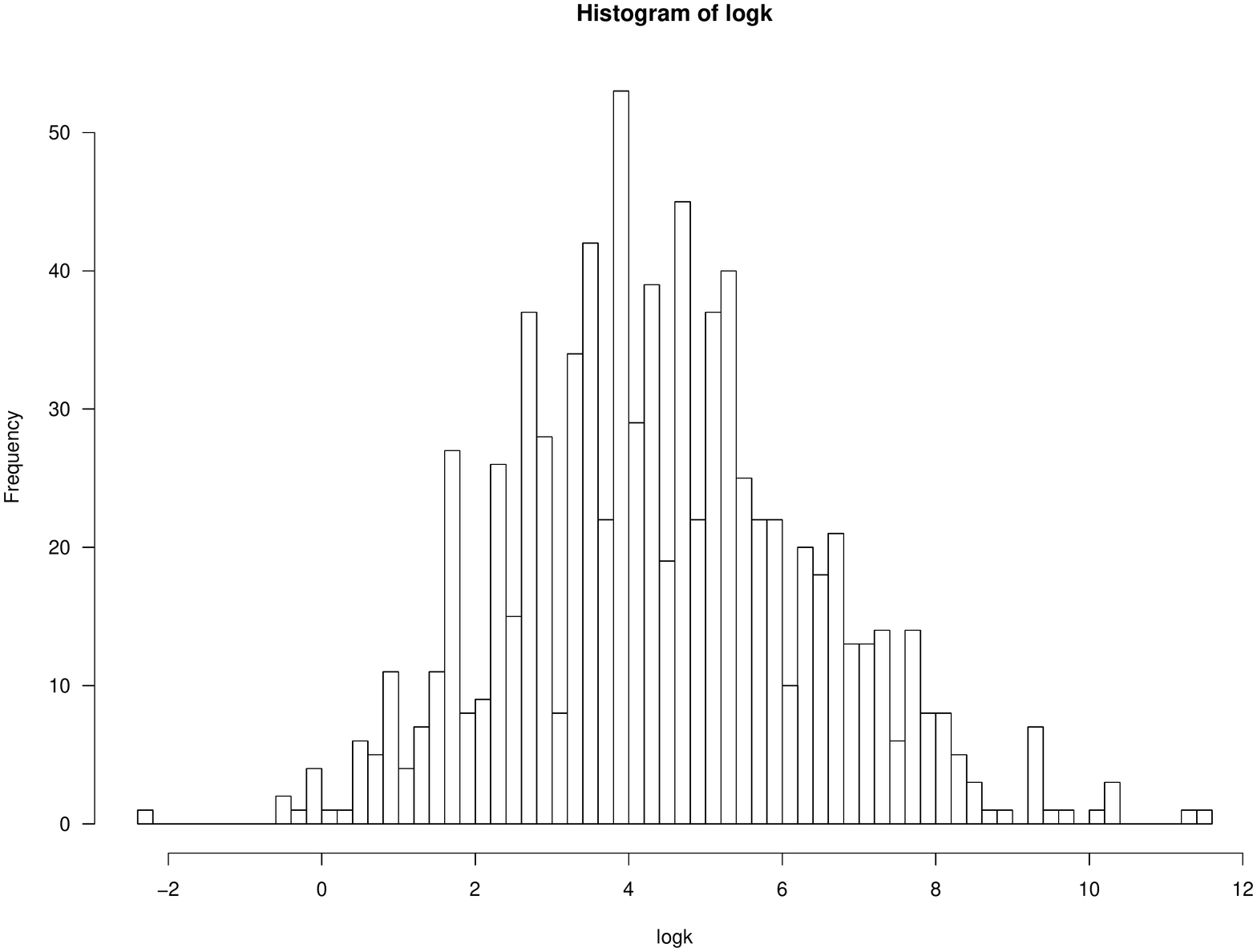}
\includegraphics[scale = .27]{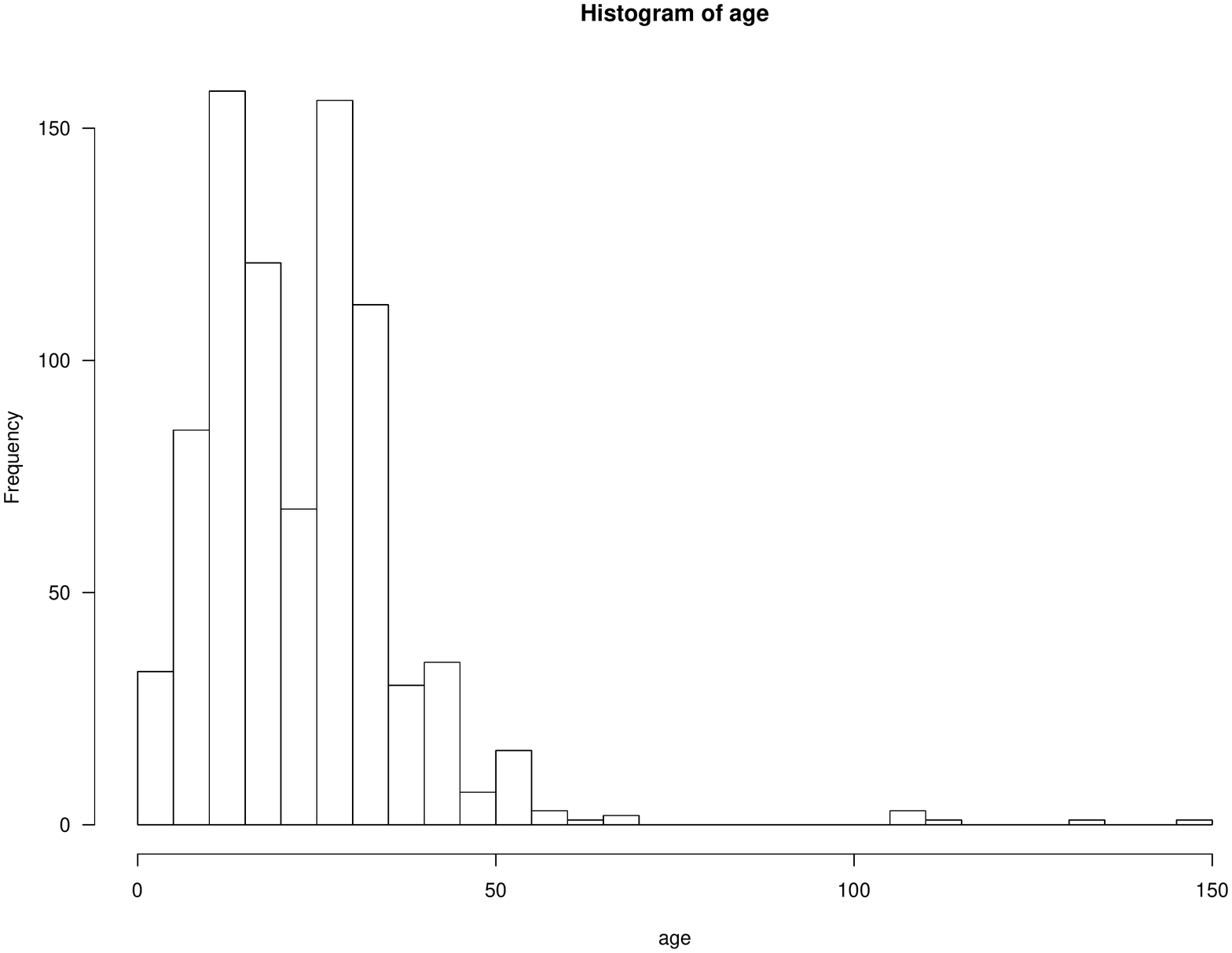}
\label{fig:capital_hist}
\end{figure}

We show in Figure \ref{fig:capital_hist} the distribution of capital managed by the firm in logs (left) and the age of the company (right). The largest company manages the capital of 97,700 million dollars. Most companies are relatively young. However, there are a few older firms that have been around for over 50 years.

\begin{figure}[!h]
\caption{Degree distribution of the co-investment network}
\centering
\includegraphics[scale = 0.3]{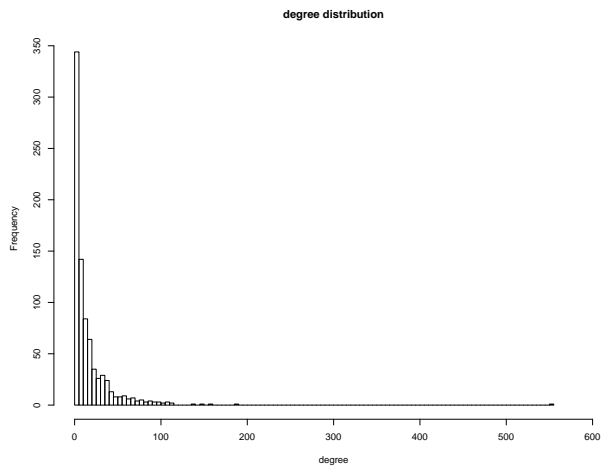}
\label{fig:degree_dist}
\end{figure}

The average firm has 34 links, while the median is 14. This substantial difference between mean and median is typical of core-periphery networks. 
In Figure \ref{fig:degree_dist} we show the whole degree distribution. The histogram shows the typical pattern of core-periphery networks; a few firms have most connections (core firms), while most companies have very few links to the core of the network (periphery firms). In the core, we find firms like Delphi Ventures, Johnson $\&$ Johnson Development Corp, and Cdib Venture Management. 

\subsection{Estimated structural parameters}
The posterior estimates for the model appear in Tables \ref{tab:endog_network} and \ref{tab:noext}. We compare the specification with endogenous network effects (Table \ref{tab:endog_network}) to a standard logit model (Table \ref{tab:noext}) that does not include any endogenous network term ($\beta=\gamma=0$). While the estimated posterior distributions are similar, below, we show that the endogenous network effects model performs better in matching crucial network properties.

In Table \ref{tab:endog_network}, the first column reports the posterior mean, while the second column is the standard deviation of the posterior. The third column shows the standard error of the posterior mean estimate. 
The fourth and fifth columns show the $95\%$ credible interval. Credible intervals represent the possible values that the model's parameters can take, given the data observed for a given significance level (usually $95\%$).

%
%\begin{table}
%\begin{tabular}{l|cccccc}
%\hline\hline
%
%Variable &  Posterior & Posterior   &  Standard & \multicolumn{2}{c}{Quantiles}\\
% &  Mean & Std. Dev.   &  Error & 2.5 & 97.5\\
%
%\hline\hline 
%Cost & -5.828 & 0.093  & 0.0009 & -6.013 & -5.647\\
%Number of partners & 0.006 & 0.0006  & 0.00006 & 0.005 & 0.008\\
%Common partners & 0.531 & 0.016 & 0.0002 & 0.499 & 0.563\\
%Same firm type & -0.025 & 0.073 &  0.0007 & -0.166 & 0.118\\
%Abs. Difference Capital (log) &  -0.022 & 0.021 &  0.0002 & -0.063 & 0.019\\
%Abs. Difference Age & -0.017 & 0.003 &  0.00003 & -0.023 & -0.011 \medskip\\
%\hline\hline
%
%\end{tabular}
%\end{table}

\begin{table}
\caption{Model with endogenous network variables}
\centering
\begin{tabular}{l|cccccc}
\hline\hline

Variable &  Posterior & Posterior   &  Standard & \multicolumn{2}{c}{Credible Interval}\\
 &  Mean & Std. Dev.   &  Error & 2.5 & 97.5\\

\hline\hline 

Cost of link ($\alpha_0$)  & -5.897 & 0.096 & 0.0001 & -6.083 & -5.709 \\
Popularity ($\beta$) & 0.006 & 0.0006 & 0.000006 & 0.005 & 0.007 \\
Common partners ($\gamma$) & 0.526 & 0.016 & 0.0001 & 0.494 & 0.558 \\
Same firm type ($\alpha_1$) & -0.054 & 0.074 & 0.0008 & -0.200 & 0.090 \\
Abs. Difference Capital (log) ($\alpha_2$) & -0.023 & 0.021 & 0.0002 & -0.064 & 0.018\\
Abs. Difference Age ($\alpha_3$)& -0.017 & 0.003 & 0.00003 & -0.023 & -0.012 \\
Same State ($\alpha_4$)& 0.641 & 0.096 & 0.003 & 0.452 & 0.829 \\
\hline\hline 

\end{tabular} \\
\scriptsize \flushleft Acceptance rate is 0.132
\label{tab:endog_network}
\end{table}

In Table \ref{tab:endog_network}, the estimated cost of forming a link is negative, as expected, and it appears relatively high, with a posterior mean of -5.987. This result implies that syndication links are costly to maintain. The $95\%$ credible interval does not include zero; therefore, we can credibly state that this estimate is negative. The estimate is precise, as shown by the posterior's small standard deviation and the posterior mean's small standard error. The economic interpretation is that forming an additional link will cost an average decrease of 5.897 in payoffs. 

The estimated cost seems relatively high, so we need to consider the benefits of partnerships. These estimates are in the rest of Table \ref{tab:endog_network}, and include both endogenous network benefits (popularity and common partners) and payoffs from exogenous attributes. The first endogenous network term is the popularity ($\beta$), which is determined by the number of 2-stars in the network. The credible interval for this parameter is positive, with a posterior mean of 0.006. Because $\beta$ is interpreted in our structural model as the marginal payoff, the economic implication is that forming a link to a partner with an additional link would increase payoff by 0.006, on average.

The second endogenous network term is the number of partners shared by the two companies ($\gamma$). This term has quite a substantial effect on the payoff; an additional shared partner would increase payoffs by an average of 0.526 (credible interval positive). This value is the marginal payoff of a shared partner. The estimate shows that shared partners have a substantial effect on payoffs, driving the clustering observed in our venture capital firms' network and explaining the aggregate core-periphery structure.

Not surprisingly, the remaining variables in Table \ref{tab:endog_network} show strong evidence of homophily for two firm attributes. Two venture capital firms belonging to the same firm type do not seem to have a higher-than-random probability of forming a link. This result is shown by the credible interval that includes zero and estimates a quite large posterior probability around zero. In a standard frequentist analysis, this would correspond to saying that the coefficient is not significant. 

One would expect that the amount of capital managed by the venture capital firm plays a vital role in determining link formation. Indeed our estimate shows that a $1\%$ increase in the difference (in absolute value) of partners' capital decreases payoffs by 0.023 on average.\footnote{The difference in capital managed by the venture capital firms is logged to facilitate interpretation.} However, the credible interval includes zero, and therefore this conclusion is weakly supported.

There is strong evidence of homophily in the age of firms. The estimated posterior for the age difference shows that a 1-year difference decreases payoffs by 0.017, providing support to age homophily. The location is critical. Payoffs increase by 0.641 on average if partnered venture capital firms are in the same state. 

In summary, venture capital firms tend to prefer syndication with firms of a different type, a similar level of managed capital, and similar age. There is a strong bias towards firms in the same state. 
However, the homophily effects for the type of firm and capital managed by the firm are not very precisely estimated, so these effects are ambiguous. 
%***HERE SHOW THAT on accounting for embeddedness, the homophily effects go away
Focusing on the structural network terms, we see a preference for syndication with firms with a higher number of partners and shared partners. This result is most likely a consequence of the reputation effect. Firms with many partners signal that they have had successful syndications in the past; firms with shared partners provide a screening device, as the joint partner can certify the quality of the previous syndications.

\begin{table}
\caption{Model with no endogenous network variables}
\centering
\begin{tabular}{l|cccccc}
\hline\hline

Variable &  Posterior & Posterior   &  Standard & \multicolumn{2}{c}{Credible Interval}\\
 &  Mean & Std. Dev.   &  Error & 2.5 & 97.5\\

\hline\hline 

Cost ($\alpha_0$) & -3.756 & 0.048 & 0.0004 & -3.849 & -3.661 \\
Same firm type ($\alpha_1$)& 0.009 & 0.041 & 0.0003 & -0.070 & 0.089\\
Abs. Difference Capital (log) ($\alpha_2$)& -0.035 & 0.012 & 0.0001 & -0.059 & -0.012 \\
Abs. Difference Age ($\alpha_3$)& -0.018 & 0.002 & 0.00002 & -0.022 & -0.014\\
Same State ($\alpha_4$)& 1.049 & 0.059 & 0.0005 & 0.933 & 1.166\\

\hline\hline

\end{tabular}\\
\scriptsize \flushleft Acceptance rate is 0.189
\label{tab:noext}
\end{table}

In Table \ref{tab:noext}, we report the posterior estimates for the standard logit model, where the specification excludes the endogenous structural network terms $\beta$ and $\gamma$. While most of the effects have a similar sign, we notice that the estimated cost is of lower magnitude; and a higher coefficient for homophily by the state. This result means that homophily imputes the equilibrium network effects for popularity and transitivity estimated in Table \ref{tab:endog_network}, rather than the clustering. Therefore, it is essential to determine which specification provides the best fit for our data because this will determine whether the observed network's core-periphery structure is generated by pure homophily effects or by the transitivity and reputation effects implied by the effect of clustering in the payoff function. Furthermore, as explained in the theory section, the effect of external economic shocks are different in a model with network effects.

\subsection{Goodness of Fit}
In the alliance literature, researchers have few ways to check the fit of a model. One of the advantages of our structural Bayesian approach is that we can check whether the estimated model posterior can replicate the network  \citep{CaimoFriel2010, Mele2010a, CaimoLomi2015, KimEtAlERGMSMJ2015}. 

We take a sample of 1000 parameter vectors from the posterior distribution estimated in the previous section to implement goodness-of-fit tests. For each of these parameter vectors, we simulate our model and generate a network. We then compare the simulated networks to the observed network. A good fit should generate networks similar to the one observed in the data, for example, in terms of similar degree distribution and similar triangle counts. 
As a robustness check, we compare our simulated networks' distribution, and the observed one for network features not included in the estimation. In our payoff specification, we explicitly include transitivity and popularity effects. Showing that our estimated model can replicate the transitivity and popularity observed in the network would not be surprising nor proof of good fit because we target these network statistics directly in our specification.

\begin{figure}
\caption{Goodness of fit tests}
\centering
\includegraphics[scale = .33]{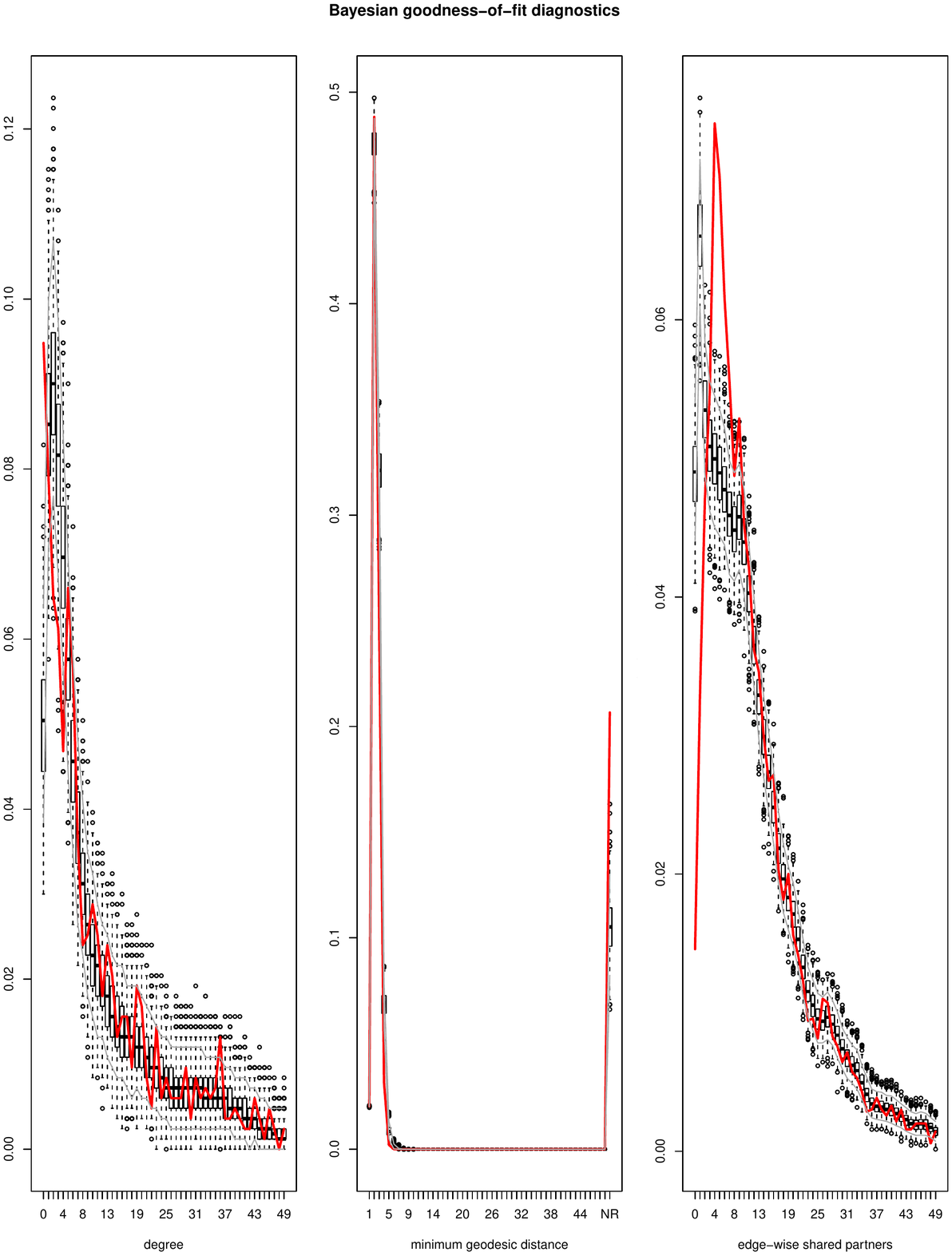}\hspace{2cm}
\includegraphics[scale = .33]{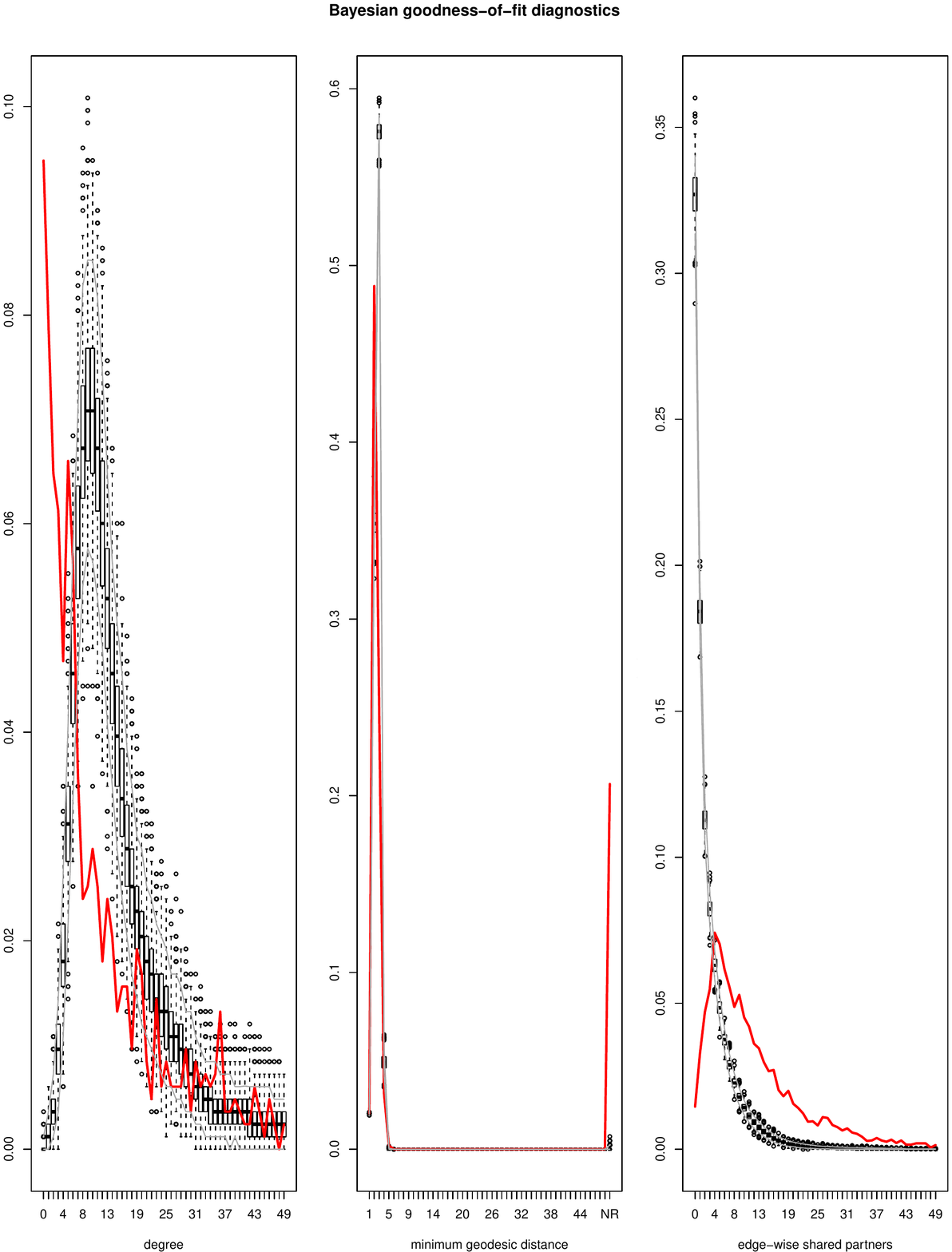}\\
Equilibrium Model (Table \ref{tab:endog_network} )\hspace{3cm} No endogenous effects (Table \ref{tab:noext})
\label{fig:gof}
\flushleft \scriptsize We report goodness of fit test for the equilibrium model estimated in Table \ref{tab:endog_network} (Left Panel) and the model without endogenous network terms in Table \ref{tab:noext} (Right Panel).  Each figure is generated through a simulation of $1000$ networks using the estimated posterior distribution. For each network, we compute degree distribution, geodesic distance, and edge-wise shared partners. The red line represents the observed network's value, while the black boxplots summarize the simulated networks. A good fit entails simulations such that the red lines fall within the boxplots limits. We conclude that our equilibrium model fits the data better.
\end{figure}

Therefore, we provide tests for network statistics that are endogenous, but not included in the payoffs. The geodesic distance is the minimum shortest path in the network between two nodes. The number of edge-wise shared partners counts the number of pairs of nodes with exactly $k$ common partners, where $k$ varies from 1 to the maximum number of nodes. Our goodness of fit test will focus on these statistics.\footnote{This is also the default output of the goodness of fit test in the R package \texttt{Bergm} that we use for our empirical analysis.}

We show the results in the Left panel of Figure \ref{fig:gof} for the model estimated in Table \ref{tab:endog_network}. This figure shows three goodness-of-fit statistics: the degree distribution, the geodesic distance, and the edge-wise shared partners. The red line represents the observed data. The boxplot is drawn at the interquartile range ($25\%$ and $75\%$). The grey lines represent $95\%$ confidence bands. If our estimated model can replicate the observed network well, then the simulated networks should have degree distributions that are "close" to the observed degree distribution. In the picture, we expect to see most of the simulations falling within the boxplots, and certainly within the $95\%$ bands. This is the case in the Left Panel of Figure \ref{fig:gof} for degree distribution, geodesic distance, and edge-wise shared partners.

We compare these test results to the goodness-of-fit test for the model estimated in Table \ref{tab:noext}. Remember that this model does not include the endogenous network statistics (stars and triangles), and it is equivalent to a logistic regression model. The fit of this specification is inferior, as shown in the Right Panel of Figure \ref{fig:gof}. This result implies that it is essential to include the endogenous network effects in the payoff functions. Therefore our equilibrium model captures essential features of the data.

%\begin{figure}
%\caption{Goodness of fit - Model without endogenous network terms - Table \ref{tab:noext}}
%\includegraphics[scale = .3]{gof_noext.pdf} \hspace{2cm}
%\includegraphics[scale = .3]{gof.pdf}
%\label{fig:gof_noext}
%\end{figure}

%\subsection{Counterfactuals Experiments}
%The main advantage of structural models is the ability to run counterfactual experiments with the estimated model, in the spirit of the Lucas Critique \textbf{(CITE LUCAS)}. When there is a policy, it is unnatural to think that the decisions of the firms will not change. Indeed our model allows us to show how firms that are optimizing their strategies will respond to such a shock in the economy. \\
%
%
%We implement a policy experiment in which the venture capital firms are allowed to form syndications only with firms in the same state. We show how this policy would impact the equilibrium networks and how the welfare of the economy changes.
%
%
%\textbf{INCREASE OF COSTs of forming syndications}

\section{Counterfactual policy experiments}
The main advantage of a structural model compared to the reduced-form models is the ability to run counterfactual experiments. For example, what would be the effect of several firms entering the market in a particular period on the equilibrium network? Answering this question using a reduced-form approach will incur in the so-called Lucas Critique \citep{Lucas1976}. This inability to run policy counterfactuals arises since a logit or probit model is a partial equilibrium model, where the aggregate equilibrium effects are not incorporated. In practice, the parameters estimated using a reduced-form approach are not policy-invariant \citep{HeckmanVytlacil2007HB}. Therefore, they cannot be used to make predictions on the effect of policy changes. On the other hand, our model includes the equilibrium feedback directly in the payoff function and models the equilibrium in the strategic network formation game.

An additional advantage of our structural model is that quantities like centrality, clustering, and network density are considered equilibrium quantities. When there is a policy change, we expect those network statistics to change as well. This change will impact how our firms make decisions about alliances and links, therefore further impacting the value of centrality, clustering, and density. This feedback effect generated by the equilibrium model is absent in a standard logit or probit model.

We use our estimated model to see how the equilibrium networks would change in three different scenarios: 1) Entry of 10 Venture capital firms in New York; 2) Entry of 5 firms in California; 3) Regulation that requires minimum capital to enter the market.\footnote{In the last counterfactual policy simulation, we assume that firms that do not fulfill the minimum capital requirement exit the market. We assume that firms do not merge to avoid the new policy's consequences because the decision to merge is not modeled explicitly in our framework. In principle, we could simulate the merging decisions using an auxiliary model, but we keep things simple in our example.}
For each policy experiment, we simulate 1000 equilibrium networks using the posterior estimates of the structural parameters.\footnote{Each network is simulated using a large number of steps, to make sure that the network is an approximate draw from the new equilibrium of the model, after the policy change.}  We compute density, clustering, homophily by firm type and state, and the degree distributions for each of these networks. We compare each of these structural equilibrium features with the observed network. This comparison allows us to determine the effect of a shock or policy change on equilibrium network architecture.

\begin{figure}
\caption{Policy counterfactual: Entry of 10 firms in New York}
\centering
\includegraphics[scale = .3]{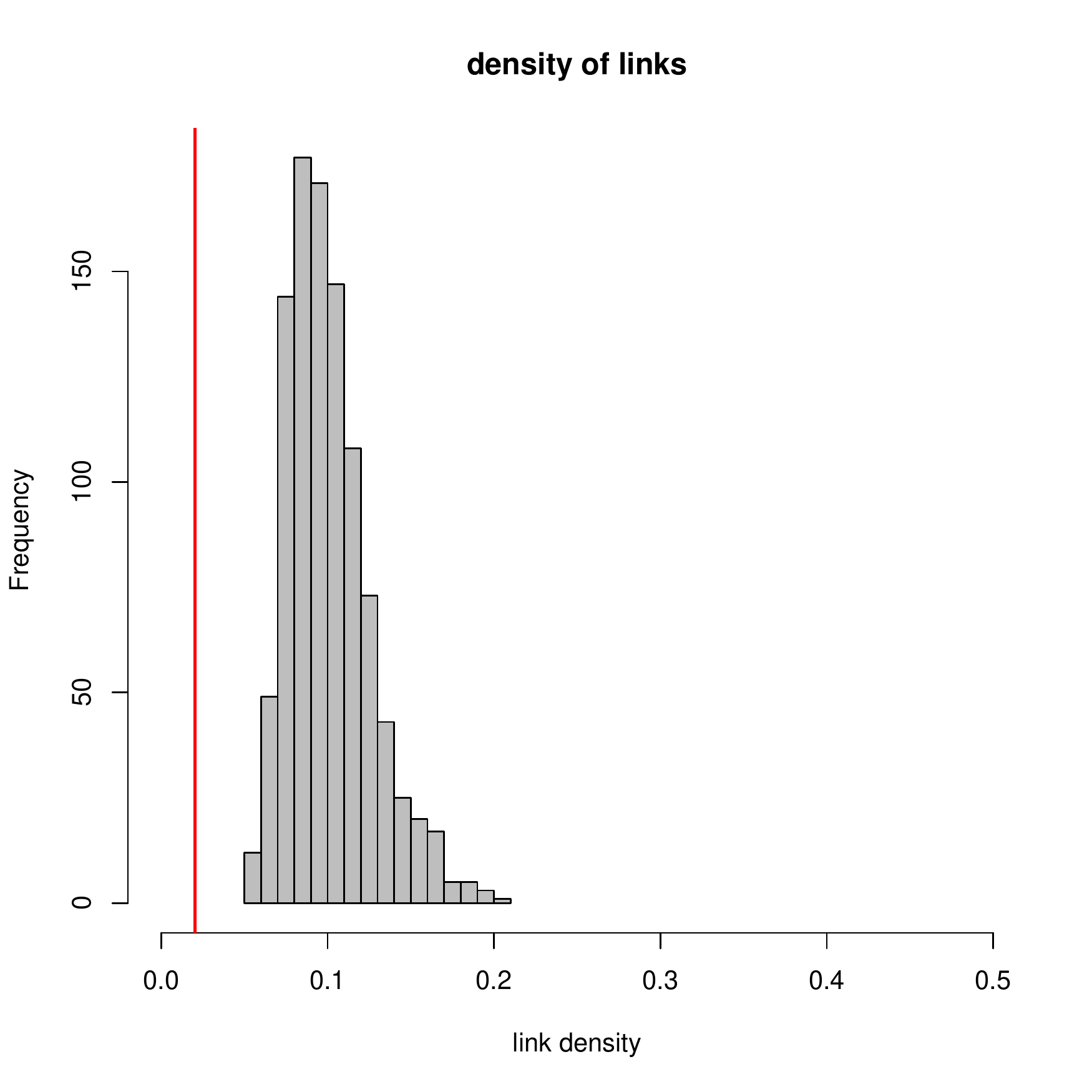}
\includegraphics[scale = .3]{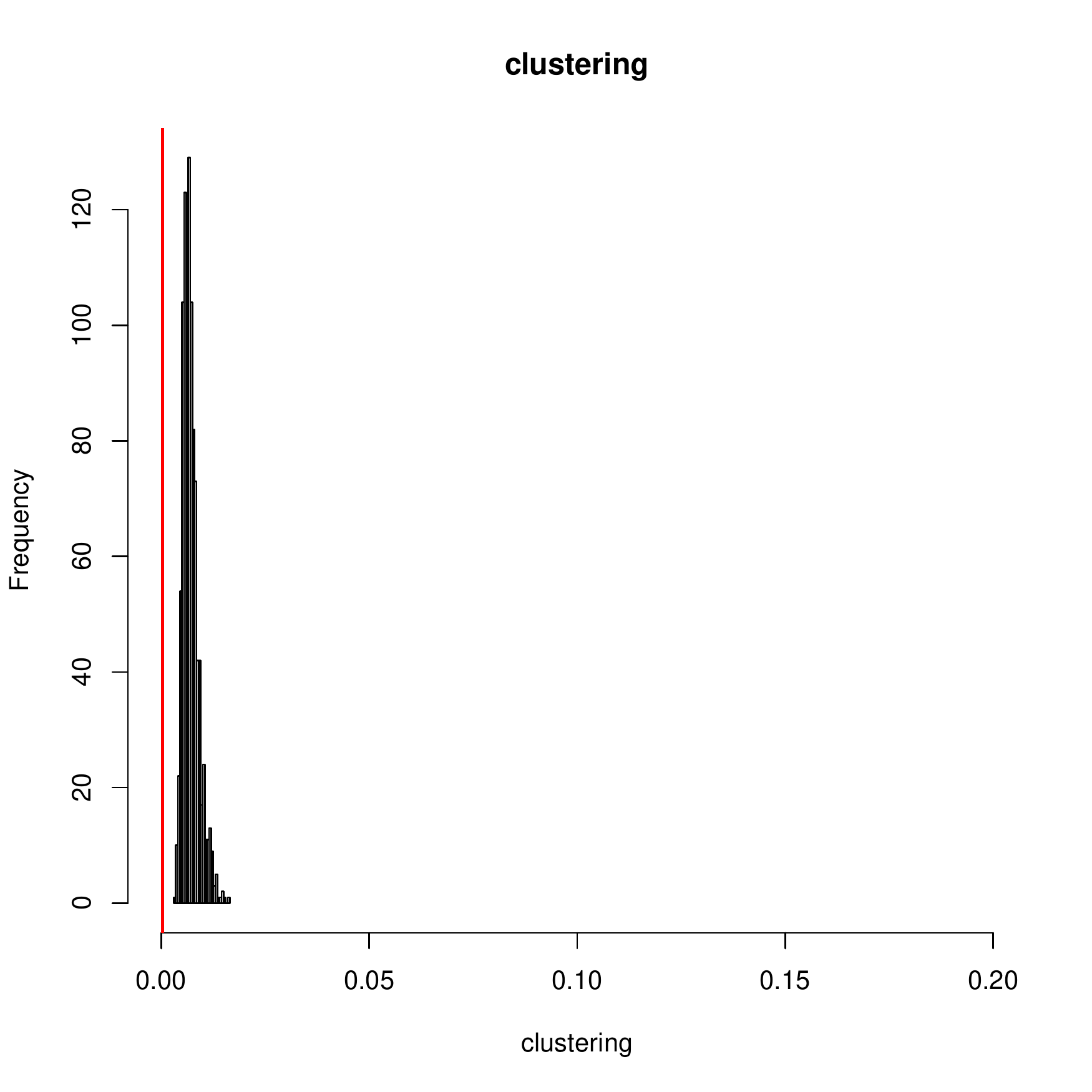}\\
(A) Link density \hspace{3cm} (B) Clustering\\
\includegraphics[scale = .3]{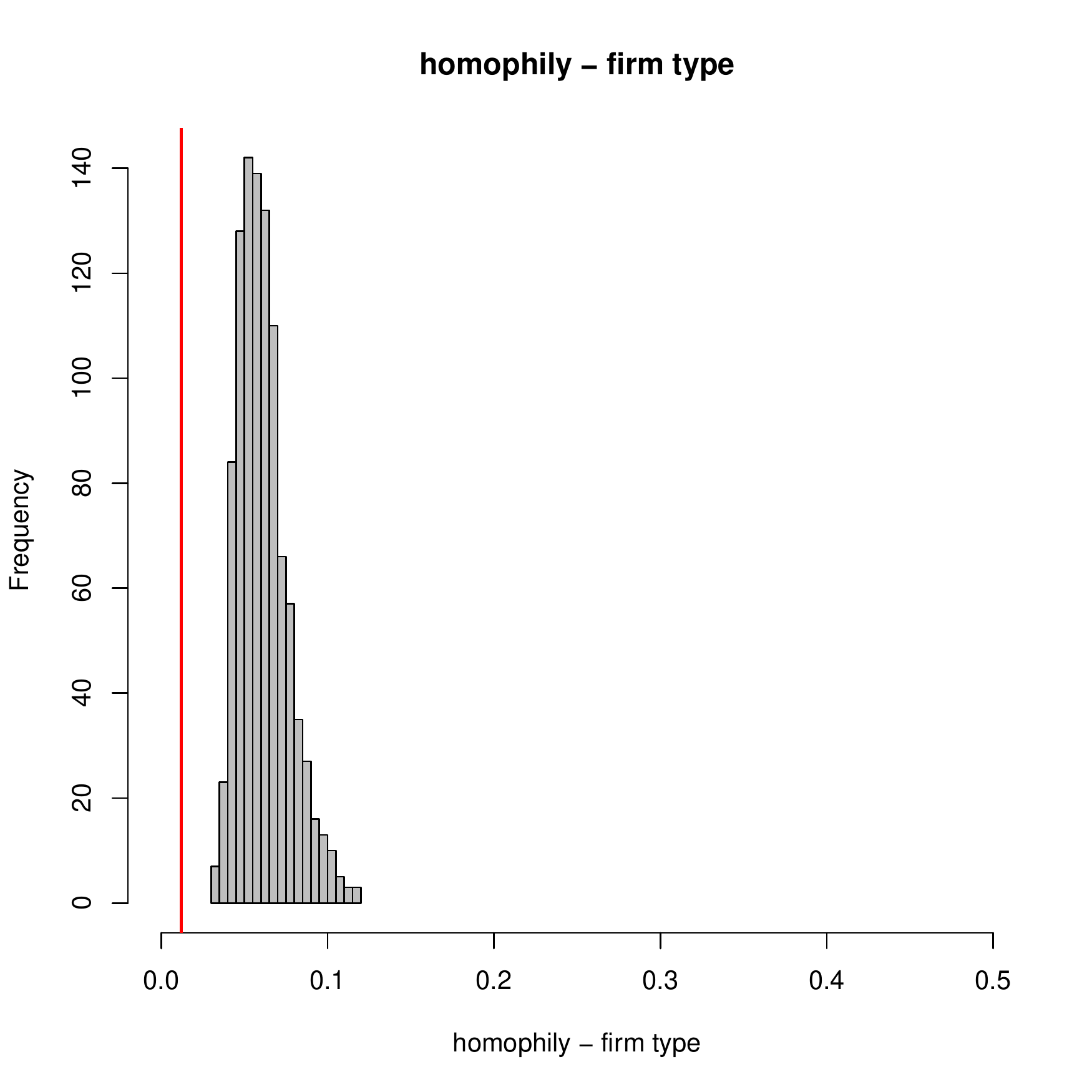}
\includegraphics[scale = .3]{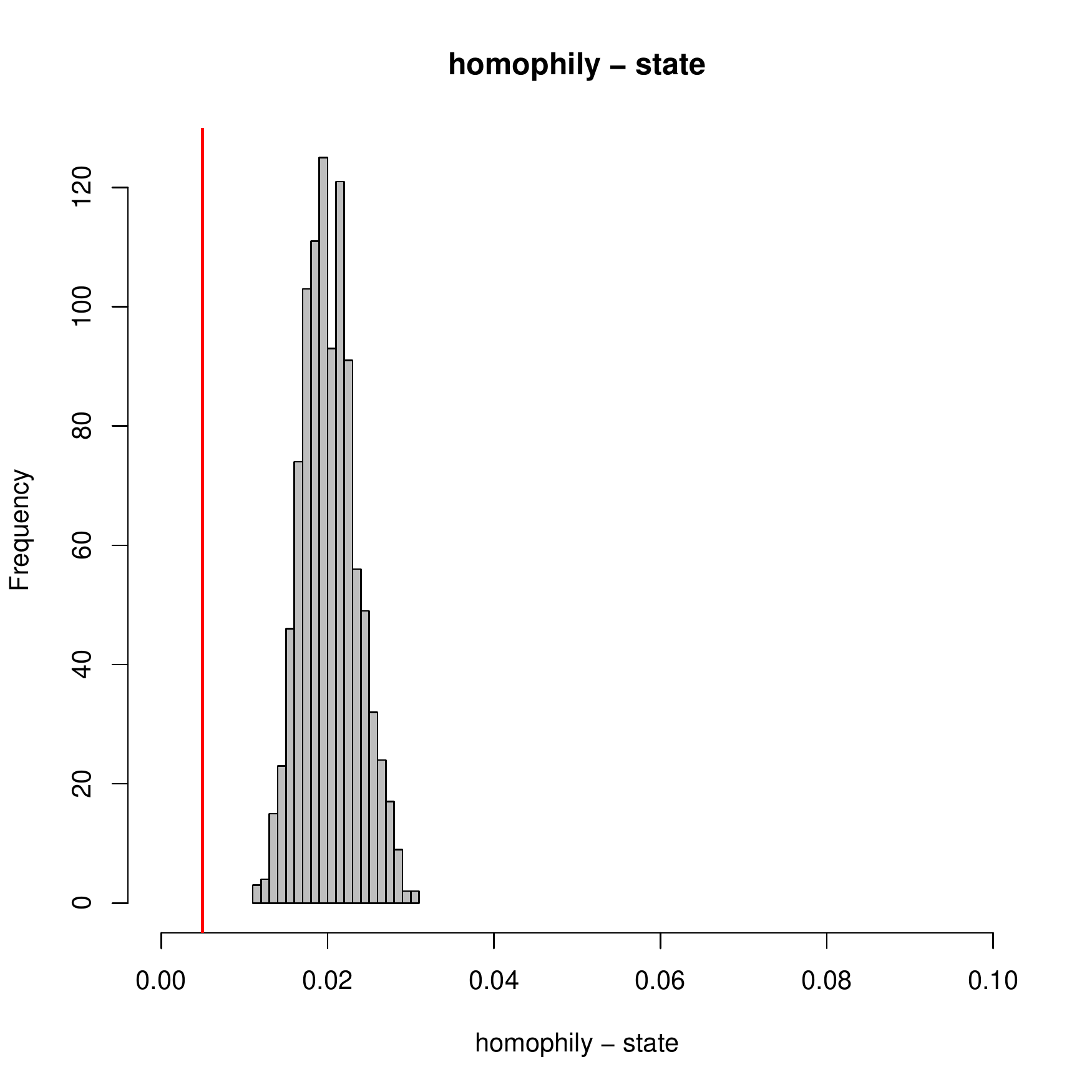}
\includegraphics[scale = .3]{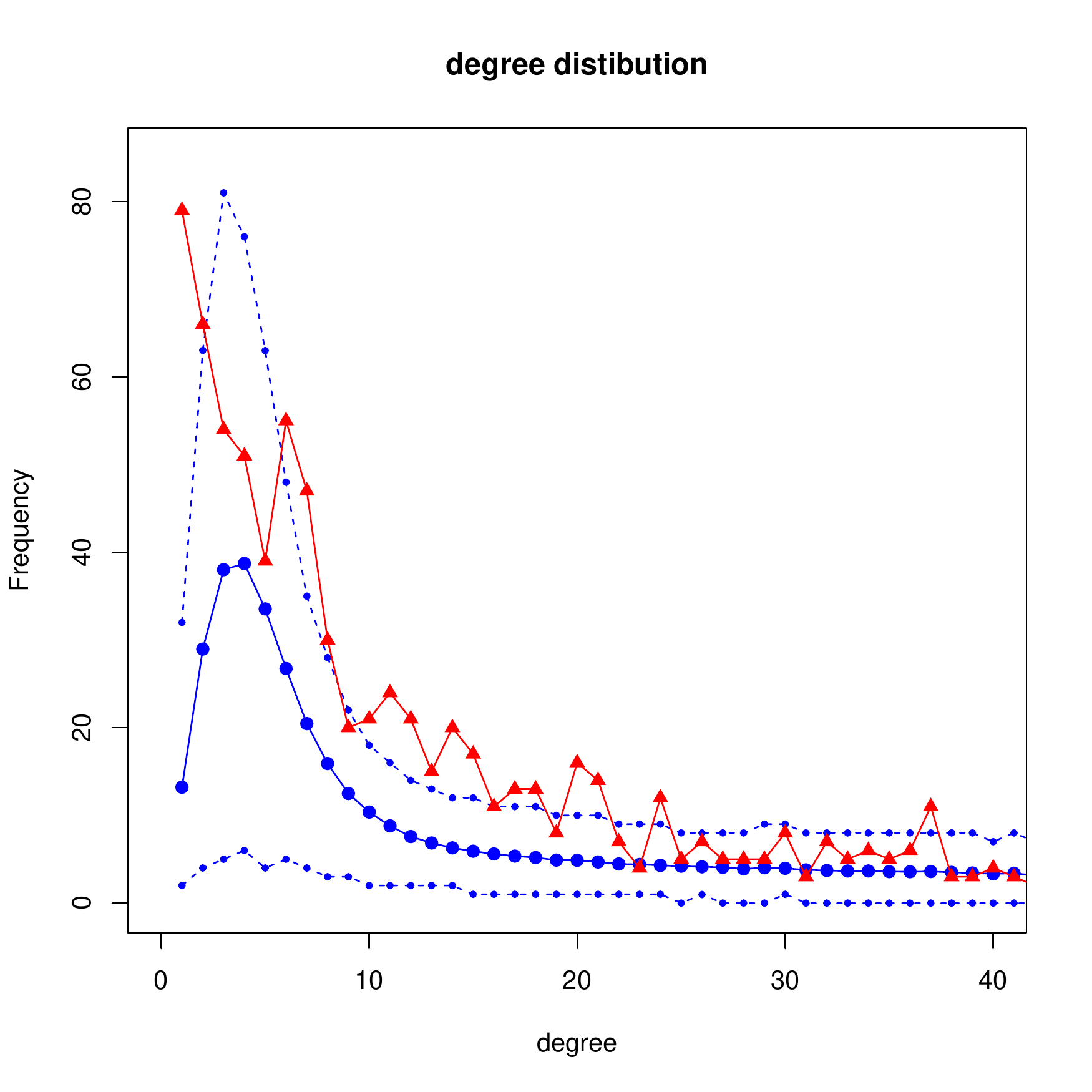}\\
(C) Homophily - Firm type \hspace{1cm} (D) Homophily - State
\hspace{1cm}(E) Degree distribution\\
\flushleft \scriptsize The red line represents the observed network feature. All the figures are obtained from 1000 network simulations from the posterior distribution. The histograms (A)-(D) show the distribution of a network feature in the counterfactual simulations. The blue lines in Panel (E) show the average degree and the  $2.5\%$ and $97.5\%$ quantiles of the degree distribution in the counterfactual simulations, while the red line is the observed degree distribution in the original network.
\label{fig:policy_10_NY}
\end{figure} 

In Figure \ref{fig:policy_10_NY}, we show the results of 1000 simulations for the entry of 10 private equity firms in the New York area. In the simulation, the (log) capital managed by the entrants is randomly drawn from a normal distribution with mean and standard deviation equal to the observed (log) capital mean and standard deviation.\footnote{This allows the entrants to be quite similar to the existing firms, in terms of capital endowments.} Therefore, the entrants are, on average, similar to the incumbent firms.

We focus on the new equilibrium network's five structural features: (A) the density of links, \footnote{We measure the ratio of the total number of links over the maximum possible number of links.} (B) the clustering,\footnote{Our measure of clustering is the total number of triangles divided by $n(n-1)(n-2)/6$. See \citep{Jackson2008} for alternative ways to measure clustering.} (C) the level of homophily by firm type, (D) the level of homophily by state, and (E) the degree distribution. In panels (A) through (D) in Figure \ref{fig:policy_10_NY}, the red lines represent the values for the network observed in our data, while the histograms show the distribution of the simulated equilibrium networks after the entry of the new firms. The blue lines in panel (E) represent the average degree distribution over 1000 simulations and the $2.5\%$ and $97.5\%$ quantiles, while the red line represents the observed degree distribution in the network data. 

Our results show that when these ten private equity firms enter the New York Market, the general equilibrium effect increases both the density and clustering of the network (on average). Furthermore, our simulations show that the network's density and clustering are significantly different from the observed network. The equilibrium level of homophily for state and firm type increases. 
The degree distribution also changes as a consequence of the new entries in the market. However, we notice that the firms used to form fewer links before the new entrants' arrival tend to form (on average) even fewer links. Firms with a low degree have high variance, which counterbalances the decrease in the number of links formed by firms with few ties. On the other hand, firms with a relatively higher degree seem to maintain their average degree. Our simulations also show a decrease in the number of isolated firms, thus contributing to the aggregate increase in the new equilibrium networks' density.

\begin{figure}
\caption{Policy counterfactual: Entry of 5 firms in California}
\centering
\includegraphics[scale = .3]{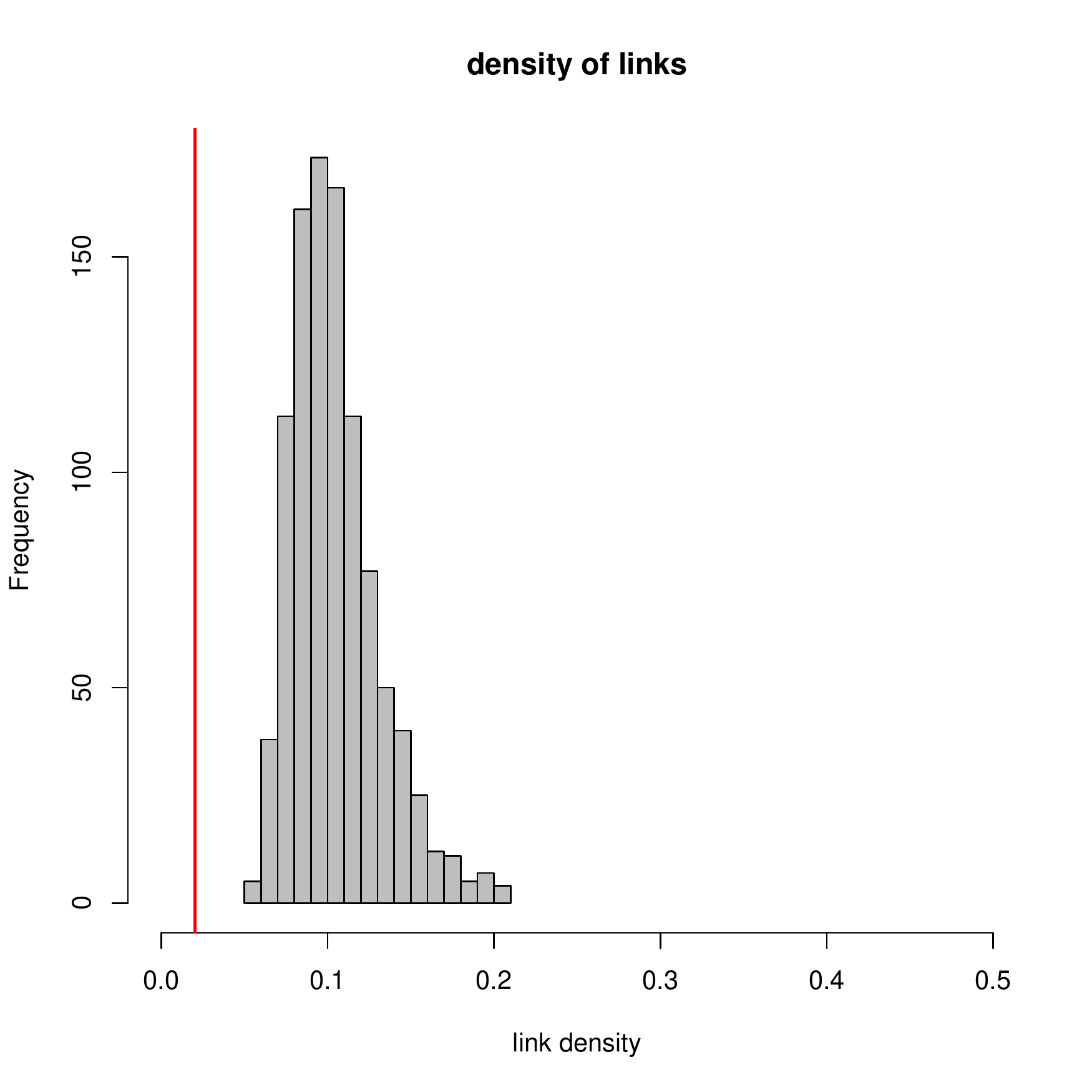}
\includegraphics[scale = .3]{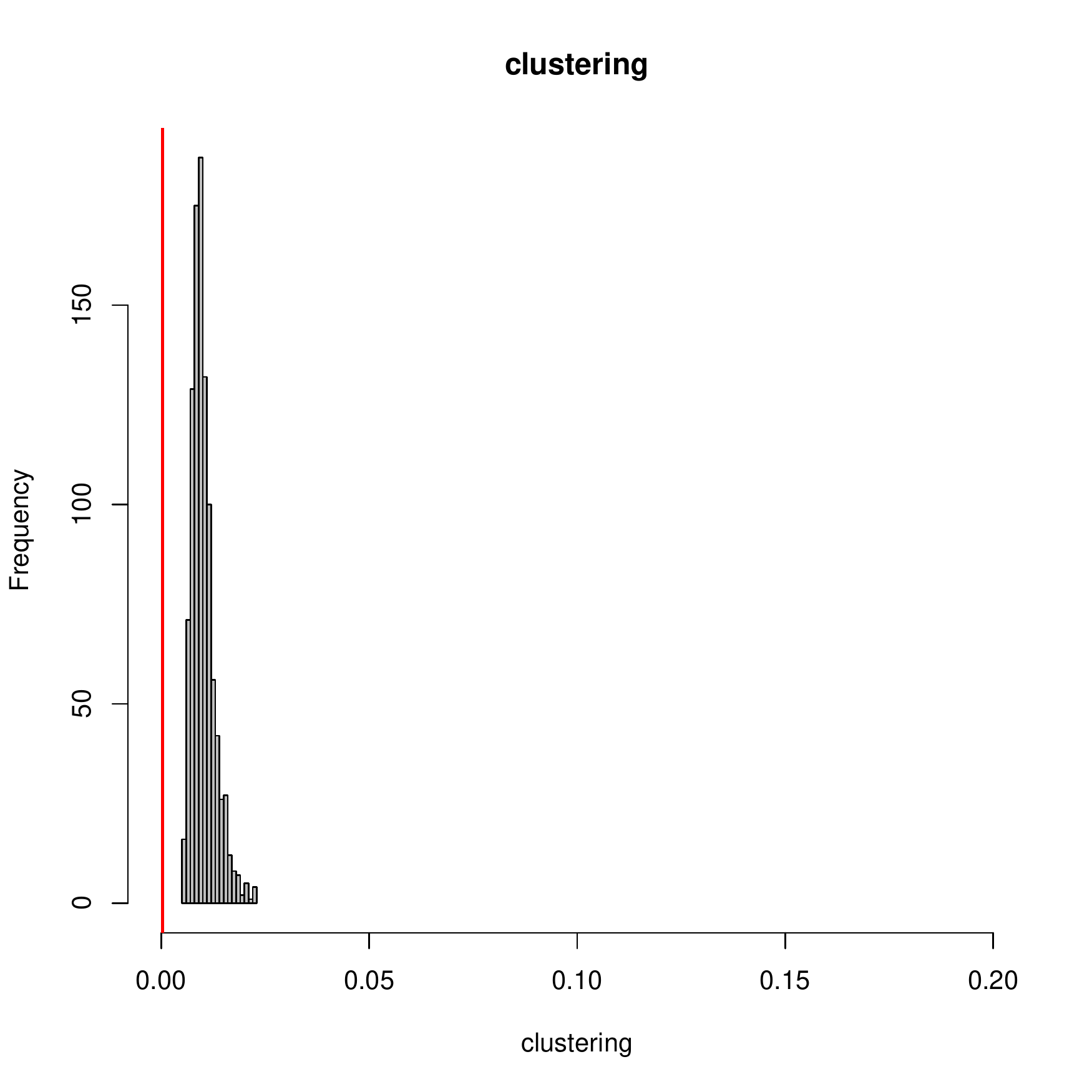}\\
(A) Link density \hspace{3cm} (B) Clustering\\
\includegraphics[scale = .3]{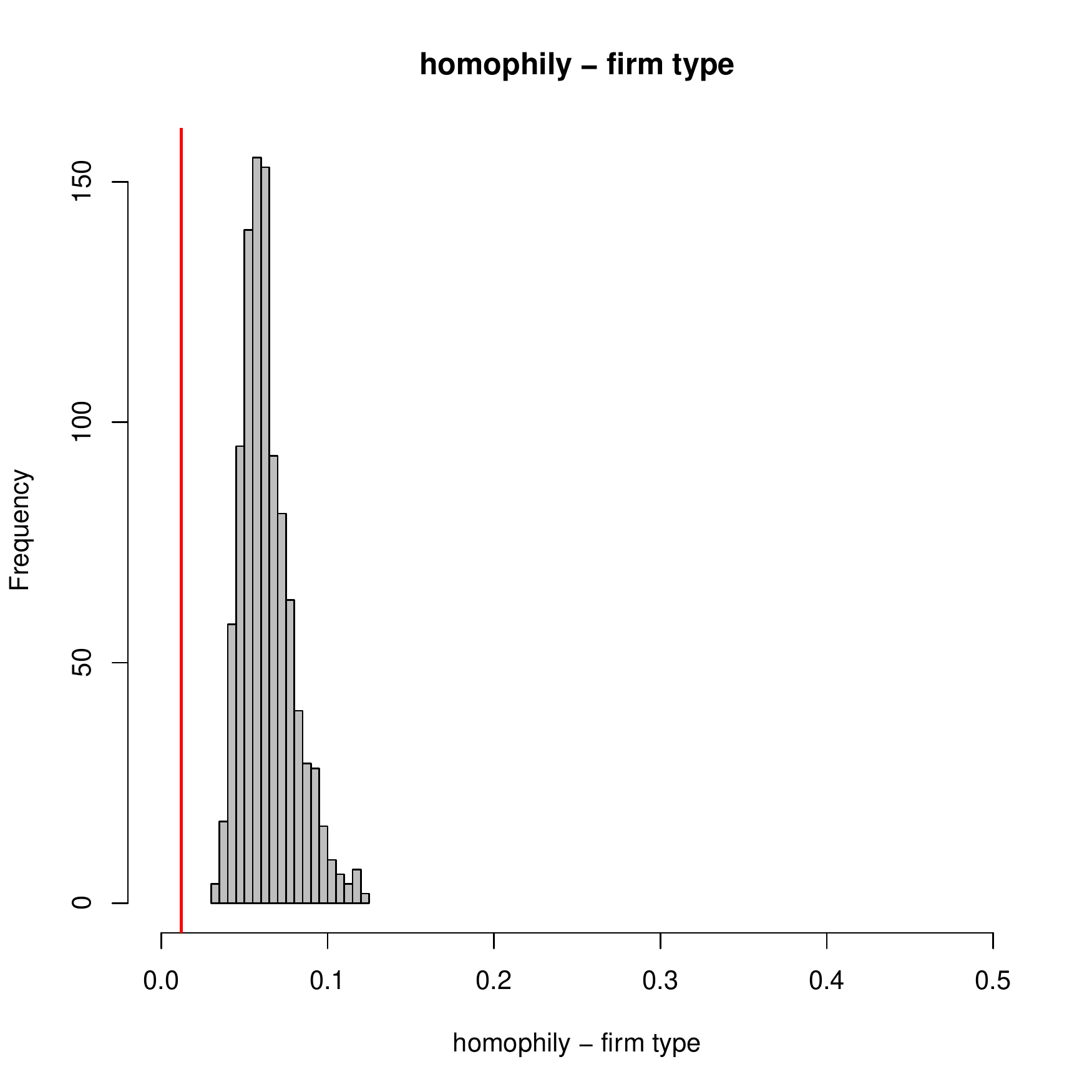}
\includegraphics[scale = .3]{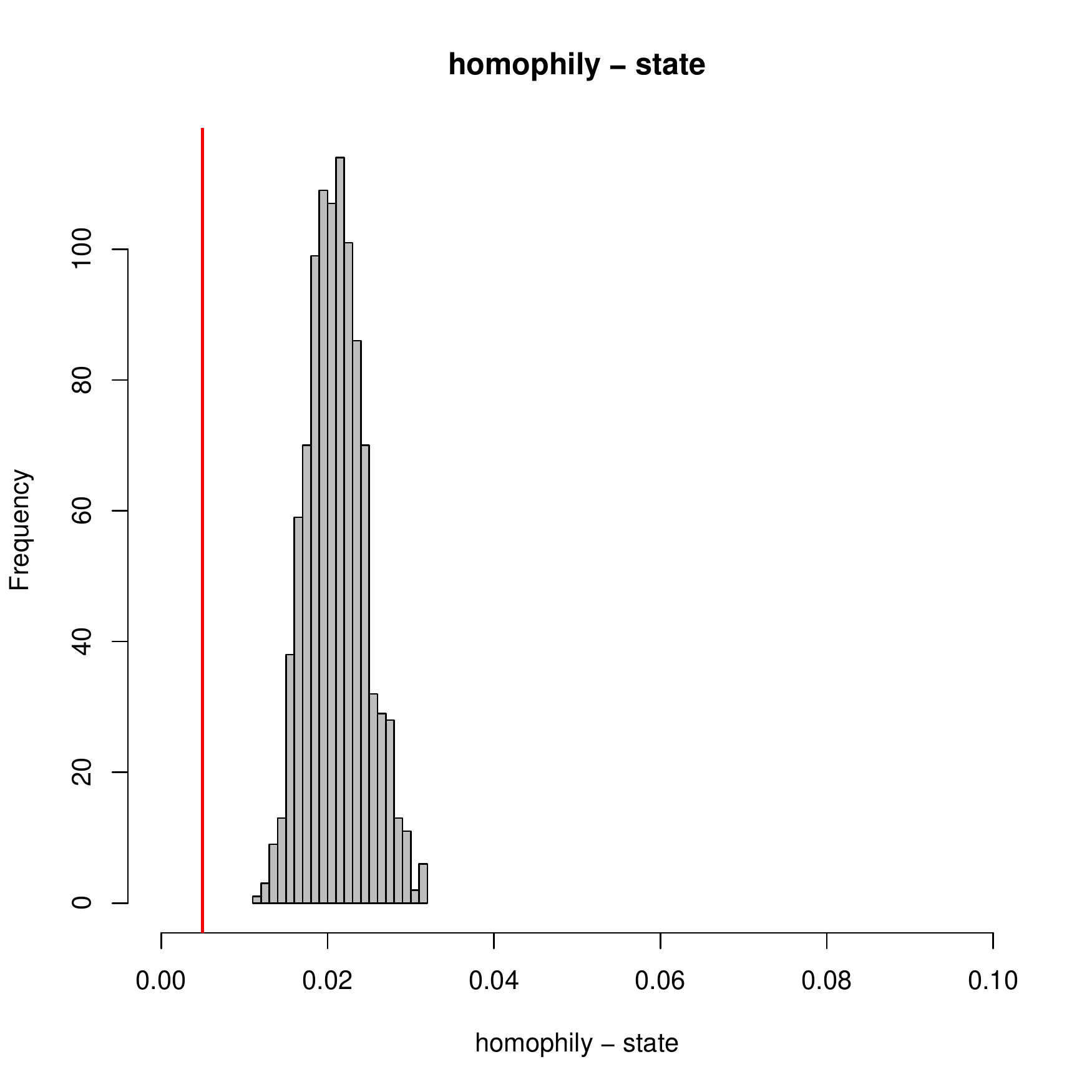}
\includegraphics[scale = .3]{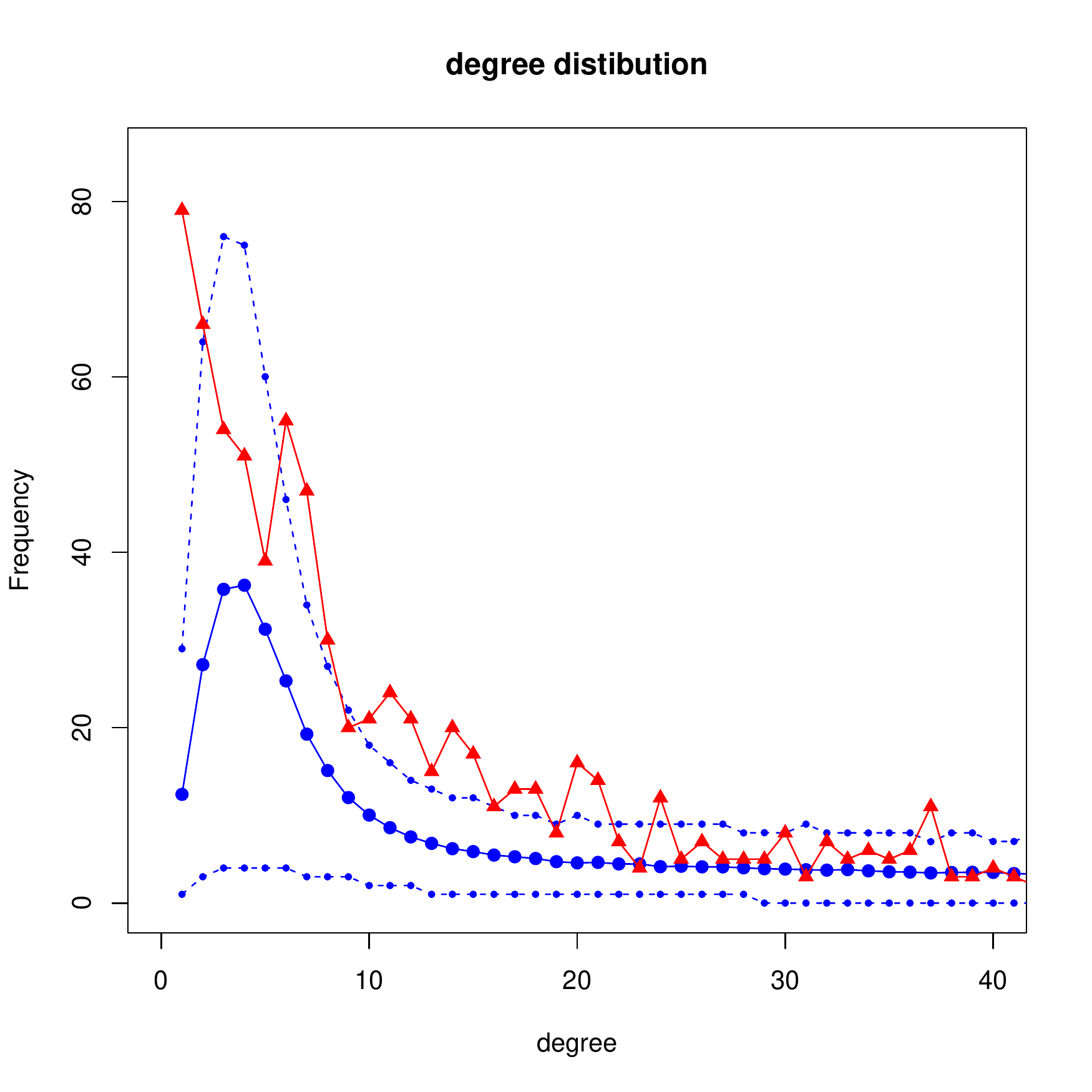}\\
(C) Homophily - Firm type \hspace{1cm} (D) Homophily - State
\hspace{1cm}
(E) Degree distribution\\
\flushleft \scriptsize The red line represents the observed network feature. All the figures are obtained from 1000 network simulations from the posterior distribution. The histograms (A)-(D) show the distribution of a network feature in the counterfactual simulations. The blue lines in Panel (E) show the average degree and the  $2.5\%$ and $97.5\%$ quantiles of the degree distribution in the counterfactual simulations, while the red line is the observed degree distribution in the original network.
\label{fig:policy_5_CA}
\end{figure}

In Figure \ref{fig:policy_5_CA}, we show a counterfactual where five venture capital firms enter the market in California.  The results are quite similar to the entry of 10 firms in New York.

The final counterfactual is a policy in which the regulator imposes a minimum capital requirement for venture capital firms. In our simulation, we impose a minimum capital that is the observed $25\%$ quantile of capital managed by the firms in the market. To keep the simulation simple, we assume that the firms that do not have enough capital will exit the market. We, therefore, abstract from the possibility of mergers to fulfill the new capital requirements. The main reason is that we have not modeled the merger decision in our model, and therefore this would not follow the spirit of the structural analysis.

\begin{figure}
\caption{Policy counterfactual: Minimum capital requirement}
\centering
\includegraphics[scale = .3]{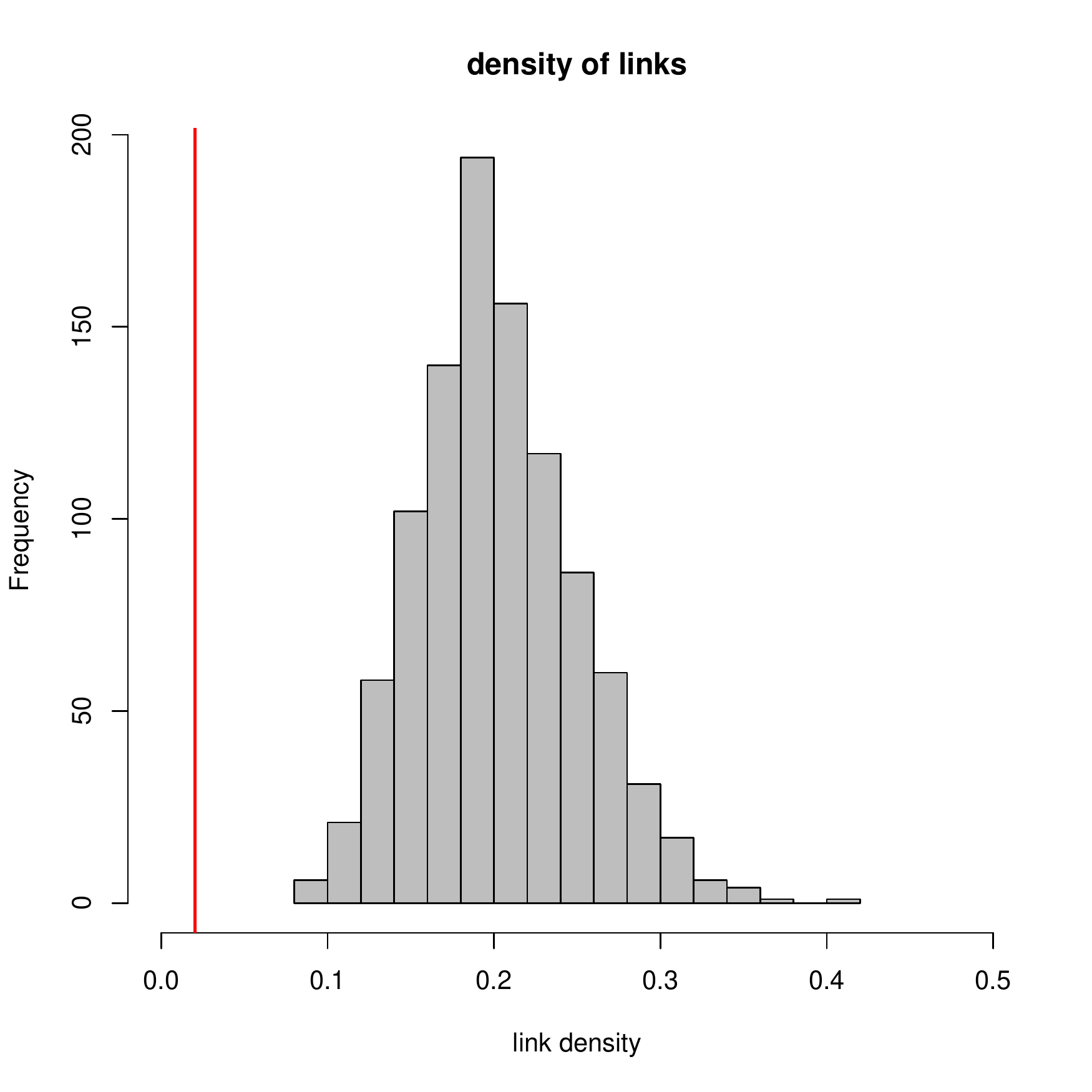}
\includegraphics[scale = .3]{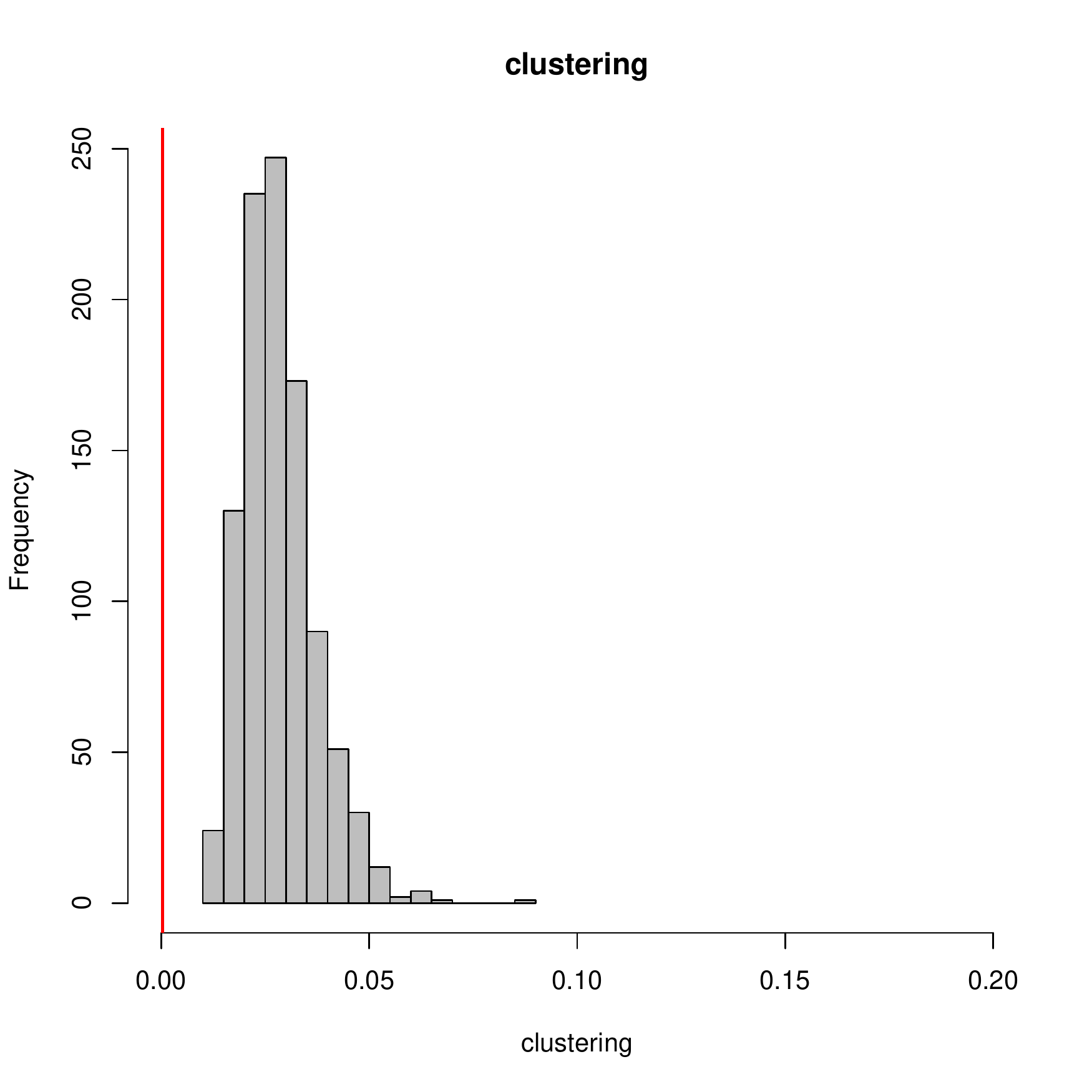}\\
(A) Link density \hspace{3cm} (B) Clustering\\
\includegraphics[scale = .3]{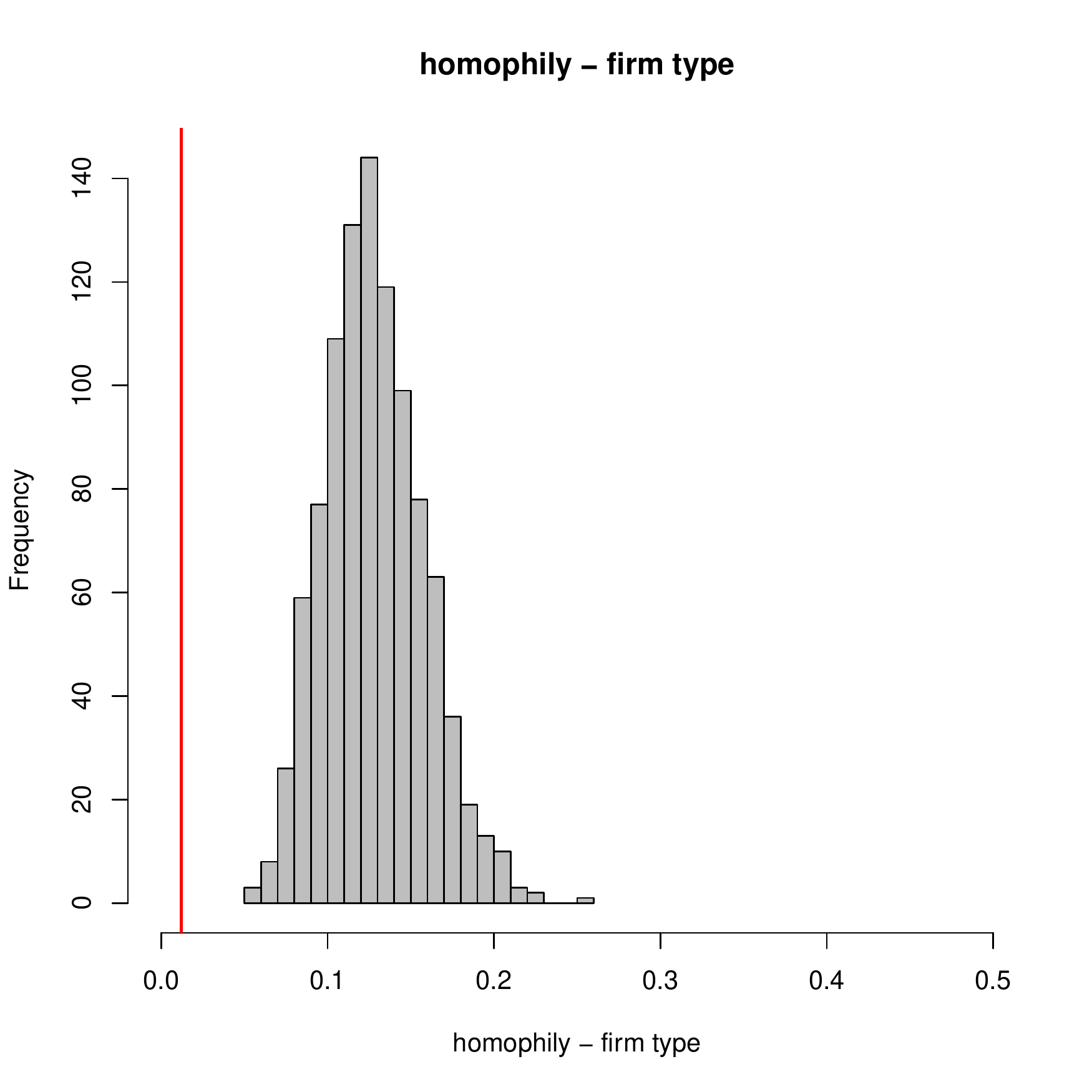}
\includegraphics[scale = .3]{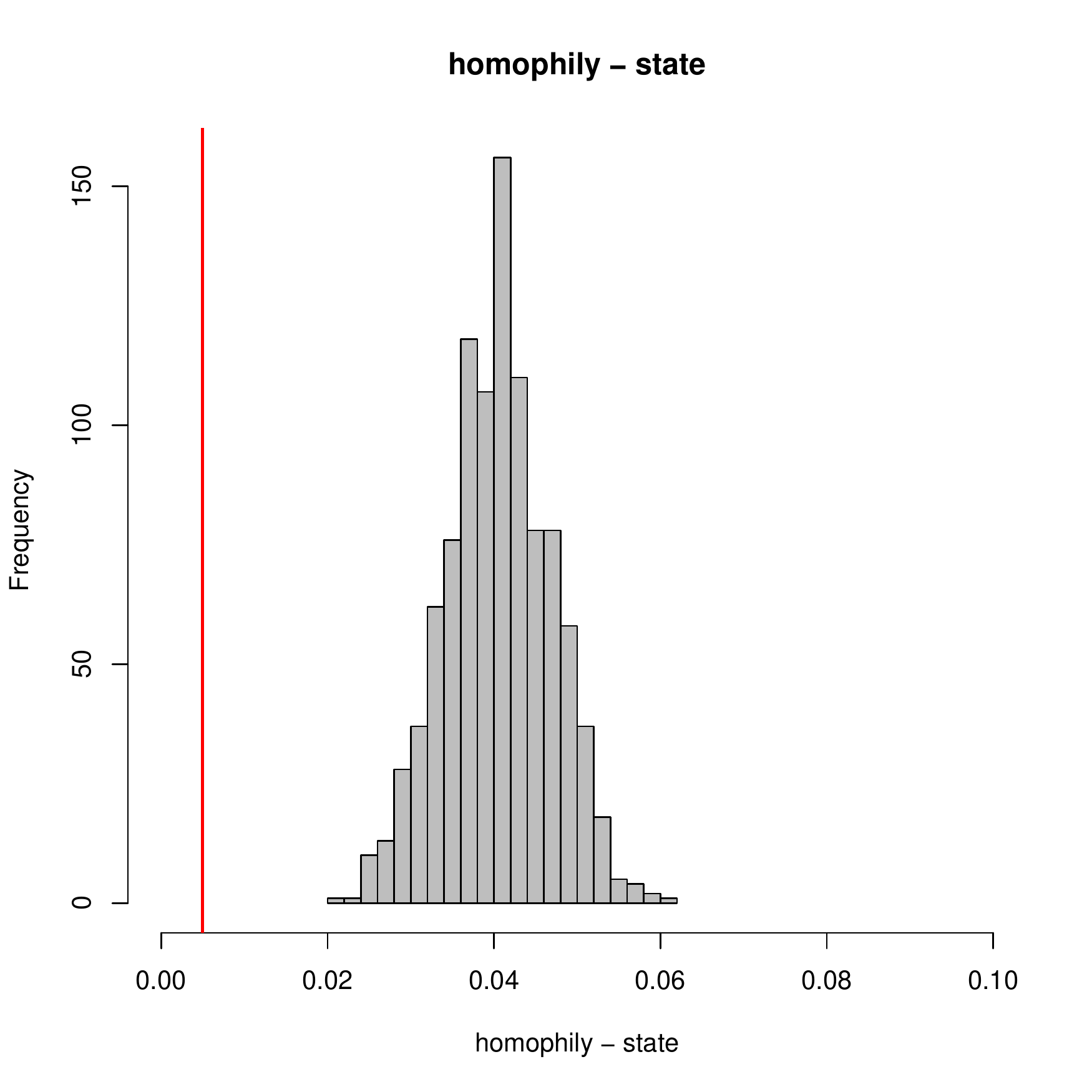}
\includegraphics[scale = .3]{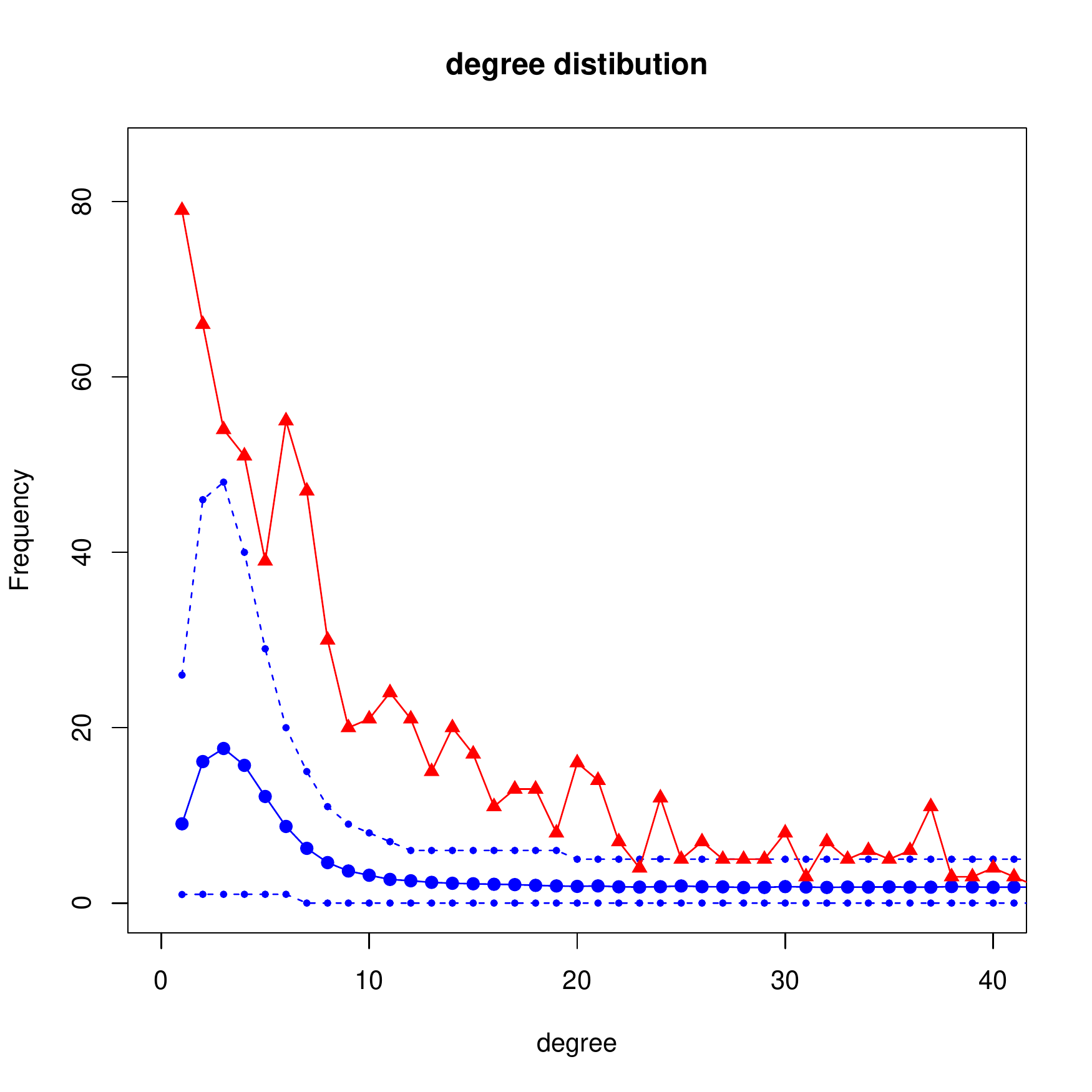}\\
(C) Homophily - Firm type \hspace{1cm} (D) Homophily - State
\hspace{1cm} (E) Degree distribution\\
\flushleft \scriptsize The red line represents the observed network feature. All the figures are obtained from 1000 network simulations from the posterior distribution. The histograms (A)-(D) show the distribution of a network feature in the counterfactual simulations. The blue lines in Panel (E) show the average degree and the  $2.5\%$ and $97.5\%$ quantiles of the degree distribution in the counterfactual simulations, while the red line is the observed degree distribution in the original network.
\label{fig:policy_minK}
\end{figure}

Figure \ref{fig:policy_minK} shows the results of the policy change counterfactual. After the policy, there are only 625 firms in the market with the minimum required capital. The new equilibrium configuration involves a denser network and a much more clustered network. These effects are much larger than in the previous counterfactuals. Also, the Panel (E) analysis shows that the degree distribution is quite different after implementing the minimum capital requirement. Most of the change is due to firms with no links before the policy change. In the new equilibrium, the number of firms with no syndication links drops on average by 70. These firms' exit contributes to the increase in density. It is not surprising that the degree distribution shifts down, as there are fewer firms in the market. As a result, the shape of the degree distribution significantly changes.

\section{Conclusions and Discussion}

Several studies have analyzed how firms establish ties to fulfill their resource needs \citep{chung2000, rothaermel2008}, access knowledge \citep{BaumCowan2010}, and gain reputation \citep{gu2014}. The most salient challenge in this research area addresses the network formation process's endogenous nature stemming from the interdependence among ties. Management scholars have empirically examined tie formation as a dyadic \cite{Ahuja2000, ahuja2009, chung2000, garcia2002, Gulati1999, li2002, rothaermel2008, hallen2008, reuer2013, stern2014} or triadic relationship \citep{davis2016} using a reduced form regression. These studies have provided valuable insight into the tie formation's micro-mechanisms and highlighted how dynamic it is. However, a reduced-form estimation using the binary choice model to understand link formation has two fundamental limitations. First, reduced form regression is lacking in its ability to account for interdependence among ties \citep{mindruta2016, KimEtAlERGMSMJ2015}.\footnote{ More recently, studies have tackled this challenge using Exponential Random Graph Models or ERGMs \citep{Snijders2002, Brennecke2020}. Despite the progress, using ERGM to estimate tie formation is that the economic interpretations based on the parameter estimates remain challenging.} Second, a reduced form regression does not enable scholars to run policy counterfactuals. This study addresses both these challenges by developing a general equilibrium model to describe the formation of ties among firms.

We develop a strategic network formation model that identifies a firm's net benefit from forming a link in equilibrium. This approach controls the endogeneity of network effects, such as transitivity, by imposing strategic equilibrium conditions on the linking decisions. Consequently, we obtain a coherent economic and statistical framework to estimate the effect of endogenous network characteristics, such as clustering or popularity. Furthermore, we show that our theoretical model's equilibrium corresponds to an exponential random graph model (ERGM), and our Bayesian estimation strategy improves existing methods by attenuating degeneracy problems \footnote{The simulation of models suffering from degeneracy problems tend to sample degenerate networks, either almost empty or almost complete. This feature of the model makes it challenging to fit the observed data and makes estimators unstable and unreliable.} and guarantees convergence to the structural parameters' posterior distribution. 

Our structural equilibrium approach's key advantage is that the estimated parameters correspond to meaningful economic concepts, such as the net marginal benefits of syndication for each firm. Additionally, the network structural features, such as density, clustering, or homophily, are strategic equilibrium quantities. This focus on the equilibrium implies that the model can predict how the structural features will change when there is an external shock such as policy change, entry, or exit of firms. Therefore, using a utility-based equilibrium model allows us to perform analysis of policy counterfactuals. These counterfactual experiments allow managers or organizations to understand the implication of their strategic decisions on the emergence of inter-organizational network structure and decode the "what if" questions.

An equilibrium model to study network formation among firms presents several advantages for organizational scholars. First, researchers and policymakers can benefit from the structural model approach to answer some crucial policy counterfactual.  An equilibrium model enables investigators to answer how firms reorganize their networks in response to a policy or economic shock -- understanding the impact of policy shock on tie formation among organizations accords academics the opportunity to be a part of policy debate. Second, for studies that use ERGM, this model provides an economic interpretation of the estimated parameters. Using the ERGM approach, a researcher can use this model as a reference to provide the economic interpretation for their parametric estimates. Our framework shows how the ERGM parameters translate into marginal costs and benefits for the firms. Furthermore, standard goodness of fit tests for the ERGM models shows that our model provides a better fit of the data than a traditional logit regression.

%\color{red}  ANGELO The model does not alleviate degeneracy, it is the bayesian estimation method that does it. I woul dtake next sentence out\color{black}Furthermore, it alleviates degeneracy issues associated with estimating ERGM models. 

%\color{red} I WOULD TAKE THIS ENTIRE PARAGRAPH OUT. YOU ARE JUST SUMMARIZING THE PAPER AGAIN... WE JUST NEED A SENTENCE IN THE PARAGRAPH ABOVE, SAYING THAT STANDARD GOF test FOR ERGMS SHOW THAT THE EQUILIBRIUM MODEL BETTER MATCHES THE NETWORK FEATURES. NOT NECESSARY TO RE-EXPLAIN THE WHOLE THING. \color{black} Furthermore, our approach has the advantage of performing tests for goodness of fit of our model. To verify our results, we check the model fit for our main specification, which includes endogenous network terms, which we compare to the fit of a model without endogenous network terms. The latter model corresponds to a logit, where links are assumed independent. The model with endogenous network terms can replicate most features of the observed data, including degree distribution and geodesic distance distribution. On the other hand, the model without endogenous network terms fails to match most of these features. We conclude that our equilibrium modeling of network formation is crucial to replicating the data's real-world features and provides an economic interpretation of the parameters. This feature of our model allows us to understand firms' incentives to form ties while controlling for endogeneity generated by the strategic equilibrium.
%

Our application to Venture Capital syndication networks provides a simple illustration of how the methodology can be implemented, from the specification of the marginal payoff functions, estimation of the model, the economic interpretation of the structural estimates, and simulation of counterfactual policy experiments.  Our results show that venture capital firms form syndicates due to their preference for firms similar to them regarding age, managed capital, and geographic location. These estimates are in line with the empirical evidence on tie formation among venture capital firms \citep{GulatiGargiulo1999, SorensonStuart2001, KogutEtAl2007}, and hold after controlling for the interdependence among ties and the equilibrium restrictions imposed by our model. The estimated parameters are interpreted as the marginal cost of each link, marginal benefit of homophily/heterophily, and marginal payoffs of popular firms and shared partners. We examine two policy counterfactuals —- how firms' entry or the imposition of a minimum level of managed capital could change the syndication network structure. Our simulations show that the network becomes more clustered and denser in response to these shocks; firms with no links before the shock will form at least a tie, thus increasing density. Most new ties are with similar firms, thus increasing the levels of aggregate homophily in the network.

While our method has several advantages, we need to acknowledge some minor limitations. First, the estimation is computationally intensive. In particular, the simulation of the posterior distribution may be more computationally intensive than the commonly used Markov Chain Monte Carlo Maximum Likelihood Estimator (MCMC-MLE). However, we note that this problem will decrease in a few years with faster hardware and better computational methods \citep{BhamidiEtAl2011, ByshkinEtAl2018, StivalaEtAl2016, MeleZhu2015}. Second, we model payoffs as functions of observable firms' characteristics, though unobserved heterogeneity may play a role in forming links. Introducing unobserved heterogeneity in our model is possible at the cost of a substantial increase in computational burden. While this is an essential direction of research, we leave the detail of this extension to future work.
    
 This approach can be adopted in management research to study other kinds of networks. Several empirical networks studied in the management literature display a similar core-periphery structure. For example, \cite{FlemingEtAl2004} examines the emergence of a dense cluster in the network of patents' co-authors in the early 1990s in Silicon Valley. This model would be applicable to study the emergence of co-author networks and predict changes in the network structure based on competition or any other policy change. Our model can be used as a framework to understand how endogenous ties create knowledge spillovers in equilibrium.

 Furthermore, our approach allows the simulation of counterfactuals, such as the effect of subsidies on research and innovation on the network structure, or how public policies affect knowledge diffusion. We believe that the equilibrium approach to networks used in this paper, combined with recent computational advances in estimation and large networks \citep{ByshkinEtAl2018, StivalaEtAl2016, MeleZhu2015}, provides a robust framework for the analysis of networks in management. We hope that our work will stimulate more researchers to apply this framework to new questions and datasets.

\newpage

%
%   or
%
 \begin{APPENDICES}

\section{Technical details for the equilibrium model of network formation}
In this appendix we provide more theoretical detail regarding the theoretical model of network formation used in the empirical analysis. 
\subsection{Setup}
The economy consists of $n$ firms and time is discrete $(t=1,2,3,...)$. A generic firm $i$ has a set of observable characteristics that we denote $x_{i}$. For example, $x_{i}$ could include the size of the firm, its industry and location. We assume that each firm has $M$ observable attributes, that is $x_{i}=\lbrace x_{i,1}, x_{i,2},...,x_{i,M} \rbrace$, and we denote as $x$ the $n\times M$ matrix that contains all the vectors of firms' observable attributes.

The network of firms is represented by an $n\times n$ adjacency matrix $g$. The element at row $i$ and column $j$ will be denoted  $g_{ij}$; we follow the convention in the literature and set $g_{ij}=1$ if there is a link between firms $i$ and $j$; otherwise $g_{ij}=0$. In our application the network is \emph{undirected} and each link requires mutual consent; therefore the adjacency matrix $g$ is symmetric (that is, $g_{ij}=g_{ji}$ for all $i,j=1,...,n$). Most of the theoretical results below can be easily extended to directed networks.\footnote{See \citep{Mele2010a} for a treatment of directed networks.}

\subsection{The network formation game}
We assume that the network is formed sequentially over time and that firms maximize the surplus generated by each link. We distinguish between \emph{opportunity} and \emph{willingness} to form a link. 

We model opportunity through a stochastic meeting process. In each period, two randomly selected firms, $i$ and $j$ can form (or delete) a link. This opportunity occurs with probability $\rho(g, x_i, x_j)$, which can depend on the existing network $g$. For example, firms with shared partners may have more frequent chances to form alliances. Furthermore, the probability $\rho$ can depend on the observable firms' characteristics $x_i$ and $x_j$; for instance, firms with similar observable characteristics may have more opportunities to form partnerships.\footnote{See \cite{CurrariniJacksonPin2009} and \cite{CurrariniJacksonPin2010} for a model where meetings are biased in favor of agents of the same group. \cite{Mele2010a}, \cite{Badev2013} and \cite{ChandrasekharJackson2012} also consider variants of this "meeting" technology.}

Upon receiving the opportunity to modify a link, firms $i$ and $j$ decide whether they want to update their connection $g_{ij}$. If the link does not exist, they decide whether to form a new link; if the link already exists, they choose to eliminate it. When choosing whether to update the link, companies behave myopically, maximizing the current surplus generated by their link.\footnote{This modeling approach has been used in previous work by \cite{Nakajima2007}, \cite{Mele2010a}, \cite{MeleZhu2015}, \cite{Badev2013}, \cite{BalaGoyal2000} and \cite{JacksonWatts2001} among others.} 

To characterize the equilibrium of the model and obtain a tractable and estimable likelihood, we make some assumptions on the payoffs and the probabilities $\rho(g, x_i, x_j)$ that govern the rate at which firms receive opportunities to create and delete links. Let $g_{-ij}$ denote the network $g$ with the exclusion of link $g_{ij}$.

\begin{assumption} \label{assumption:meeting} The meeting process is i.i.d. over time, the probability that firms $i$ and $j$ meet is 
\begin{equation}
\rho(g, x_i, x_j) = \rho(g_{-ij}, x_i, x_j)>0
\end{equation}
and the sum of these probabilities over all possible pairs of firms is one.
\end{assumption}

Assumption \ref{assumption:meeting} guarantees that any pair of firms have an opportunity to update their links. The probability $\rho(g_{-ij}, x_i, x_j)$ can be very small, but it is necessarily positive. The main implication is that the network formation process can reach any equilibrium network with positive probability (\cite{Mele2010a}). The simplest probability model for $\rho$ that satisfy Assumption \ref{assumption:meeting} is a discrete uniform distribution. More generally, the probability $\rho$ can depend on the network $g_{-ij}$. The crucial part of Assumption \ref{assumption:meeting} is that the probability for any pair is positive. 

Firms' payoffs are across the networks, $g$, and observable characteristics $x$.
We will denote as $U_{i}(g,x;\theta)$ the payoff of firm $i$ from network $g$, observable characteristics $x$, and parameters
 $\theta = \lbrace\alpha,\beta,\gamma\rbrace$.

\begin{assumption} \label{assumption:utilityscalar}
The payoff of firm $i$ is
\begin{equation}
U_{i}(g,x;\theta) = \sum_{j=1}^{n}g_{ij}\left[ u(x_i,x_j;\alpha) + \beta \sum_{r\neq i,j}^{n} g_{jr} +
\gamma \sum_{r\neq i,j}^{n} g_{jr}g_{ri}  \right]
\end{equation}
where $\alpha$, $\beta$ and $\gamma$ are parameters.
\end{assumption}
The payoff $U_{i}(g,x;\theta)$ has three components. First, when firm $i$ forms a link with firm $j$, it receives a net payoff $u(x_i,x_j;\alpha) $ that depends on characteristics $x_i$ and $x_j$, and a parameter $\alpha$. For example, a firm may find companies in the same industry more attractive for a partnership. 
The payoff $u(x_i,x_j;\alpha) $ includes both \emph{costs and benefits} of direct connections, so it should be interpreted as \emph{net direct benefit} of forming a link. We follow the convention in the strategic network literature and assume that \emph{firms pay a cost for direct links, but that indirect connections are free} (\cite{JacksonWolinski1996}, \cite{Jackson2008}).
In the empirical section we provide an explicit functional form for $u(x_i,x_j;\alpha) $.

Second, when firm $i$ connects to a firm $j$, it receives an additional payoff $\beta$ for each firm that has formed a link to $j$ in previous periods. If firm $j$ has 2 links, then firm $i$ will receive $2\beta$; if firm $j$ has 5 links, then firm $i$ will receive $5\beta$. If $\beta>0$, firm $i$ prefers to link to a "popular" firm; vice-versa, there could be a competition effect, and $i$ may receive less utility from a "popular" firm since this firm has to share resources with other partners. The sign and magnitude of $\beta$ is ultimately an empirical question. Third, firm $i$ receives a payoff of $\gamma$ for each partner in common with $j$. The term $\sum_{r\neq i,j}^{n} g_{jr}g_{ri}$ corresponds to the number of common partners between $i$ and $j$.

This part of the payoff captures the clustering effects or triadic closure process. When $\gamma$ is positive, a firm receives more surplus from companies with which it shares many partners. Viceversa, when $\gamma$ is negative, a firm receives a negative surplus when linking to a company with many shared partners. In the network literature, there is an empirical regularity: if two nodes have a common neighbor, there is a high chance that they form a link (\cite{WassermanFaust1994}, \cite{WassermanPattison1996}, \cite{Jackson2008}). This model accounts for this property through the payoff structure, generating the triadic closure property as an equilibrium outcome. On the other hand, we do not impose a positive parameter $\gamma$ and instead let the data determine the sign. Therefore our model can accommodate both transitivity and intransitivity. 

Finally, we assume that firms receive a joint matching shock $\varepsilon_{ij}=(\varepsilon_{0,ij}, \varepsilon_{1,ij})$ before choosing whether to update a link. The random shock models idiosyncratic reasons that could affect the decision to link. For example, in some periods, two firms may be a  bad match (negative matching shock) for reasons that are unobservable to the researcher, such as misaligned long-term strategies or incompatible risk profiles. On the other hand, there are periods in which those companies may be a good match (positive matching shock), increasing both firms' willingness to create a partnership.

\begin{assumption}\label{assumption:shocks}
Firms receive a logistic matching shock before updating their links, which is i.i.d. over time and across pairs.
\end{assumption}

Assumption \ref{assumption:shocks} is standard in discrete choice models and random utility models in the empirical literature. In our model, this assumption is crucial to derive the likelihood of the network in closed-form,\footnote{See also \cite{Mele2010a}, \cite{MeleZhu2015}, \cite{ChandrasekharJackson2012} and \cite{Heckman1978}.} allowing us to perform maximum likelihood estimation or Bayesian estimation.\footnote{As an alternative to assumptions 1-3, we could use the spatial GMM (\cite{Conley1999}, \cite{ConleyTopa2007}) or the Approximate Bayesian Computation (ABC). These techniques do not require knowing the likelihood in closed-form (\cite{MarjoramEtal2003}, \cite{Koenig2016}).} 

\subsection{Characterization of Equilibrium Networks}
We focus on equilibrium networks that satisfy \emph{pairwise stability with transfers}, one of the most common equilibrium notions used in the network literature in economics.\footnote{See \cite{Jackson2008}, \cite{MeleZhu2015} and  \cite{ChandrasekharJackson2012} for examples.}
This equilibrium notion requires that both firms consent to form a new link. However, the surplus generated by a new link is allowed to differ between the partners because firms can shift part of the other party's payoff. This is a way to model asymmetric bargaining power or different investments of resources in the partnership. 

We will denote the transfer from firm $i$ to firm $j$ as $\tau_{ij}$. By definition, $\tau_{ij}=-\tau_{ji}$. 
Conditional on being randomly selected, firms $i$ and $j$ will form (or keep) a link if the surplus generated by forming the link (including transfers and matching shocks) is larger than the surplus without the links, that is if
\begin{eqnarray}
U_i (g_{ij}=1, g_{-ij}, x_i,x_j;\theta) + U_j (g_{ij}=1, g_{-ij},x_j,x_i;\theta) + \varepsilon_{1,ij} \geq \notag \\ 
 U_i (g_{ij}=0, g_{-ij},x_i,x_j;\theta) + U_j (g_{ij}=0, g_{-ij},x_j,x_i;\theta) + \varepsilon_{0,ij}
 \label{eq:transfer_cancels}
\end{eqnarray}
In equation (\ref{eq:transfer_cancels}) the transfers do not appear; if $i$ tranfers a positive amount $\tau_{ij}$ to $j$, then $i$'s payoff will be $U_i (g_{ij}=1, g_{-ij}, x_i,x_j;\theta) - \tau_{ij}$ while for $j$ the payoff will be $U_j (g_{ij}=1, g_{-ij},x_j,x_i;\theta) + \tau_{ij}$. When we sum the payoffs to compute the total surplus of the link, the transfers cancel out. Therefore, the condition (\ref{eq:transfer_cancels}) can be re-stated as follows
\begin{eqnarray*}
U_i (g_{ij}=1, g_{-ij}, x_i,x_j;\theta) + U_j (g_{ij}=1, g_{-ij},x_j,x_i;\theta)  - \\
  \left[U_i (g_{ij}=0, g_{-ij},x_i,x_j;\theta) + U_j (g_{ij}=0, g_{-ij},x_j,x_i;\theta)\right] \geq  \\
  \varepsilon_{0,ij}-  \varepsilon_{1,ij}
\end{eqnarray*}
Let's define 
\begin{eqnarray*}
\Delta_{ij} := U_i (g_{ij}=1, g_{-ij}, x_i,x_j;\theta) + U_j (g_{ij}=1, g_{-ij},x_j,x_i;\theta)  - \\
  \left[U_i (g_{ij}=0, g_{-ij},x_i,x_j;\theta) + U_j (g_{ij}=0, g_{-ij},x_j,x_i;\theta)\right]
  \end{eqnarray*}
Our Assumption \ref{assumption:shocks} implies that the matching shocks are logistic. Therefore the difference of matching shocks $  \varepsilon_{0,ij}-  \varepsilon_{1,ij}$ is also logistic. The main consequence is that the probability that firms $i$ and $j$ form (or keep) a link in any period has the standard logit form (\cite{Heckman1978})
\begin{equation}
P(g_{ij}=1\vert g_{-ij}, x_i,x_j,\theta) = \frac{\exp\left[\Delta_{ij} \right]}{1+\exp\left[\Delta_{ij} \right]}
\label{eq:ccp_ergm}
\end{equation}
Equation (\ref{eq:ccp_ergm}) is a \emph{conditional} probability. It describes the probability of a link between $i$ and $j$; given the rest of the network's links $g_{-ij}$, the observable characteristics $x_i$ and $x_j$, and the parameters of the model $\theta$. This probability is also implicitly conditioned on the event that firms $i$ and $j$ have an opportunity to revise their link, which happens with probability $\rho(g_{-ij}, x_i, x_j)$. The probability (\ref{eq:ccp_ergm}) shows how the link $g_{ij}$ depends on the other links in the network $g_{-ij}$, thus relaxing the independence assumption implicit in a standard logit model. Indeed, this is what strategic network formation implies: there is dependence among linking decisions, induced by the strategic equilibrium.

The model generates a sequence of networks as a result of link creation or deletion. This sequence is Markovian and converges to a unique stationary distribution that characterizes the probability of observing a specific network architecture in the long run, as shown in \cite{Mele2010a}.

Let's define the aggregate function
\begin{equation}
Q(g,x;\theta) = \sum_{i=1}^{n}\sum_{j=1}^{n}g_{ij}u(x_i,x_j;\alpha) + \frac{\beta}{2}\sum_{i=1}^{n}\sum_{j=1}^{n}\sum_{r\neq i,j}^{n}g_{ij}g_{jr} +
 \frac{2\gamma}{3}\sum_{i=1}^{n}\sum_{j=1}^{n}\sum_{r\neq i,j}^{n}g_{ij}g_{jr}g_{ri}
\label{eq:potential}
\end{equation}

Equation (\ref{eq:potential}) is called a \emph{potential function}, and it summarizes the incentives of each firm to form links, net of the matching shock. The crucial property of the potential is
\begin{equation}
 Q(g,x;\theta) - Q(g^{\prime},x;\theta) =
U_i(g,x;\theta)+U_j(g,x;\theta) - \left[U_i(g^{\prime},x;\theta)+U_j(g^{\prime},x;\theta)\right]
\label{eq:potential_properties}
\end{equation}
where $g$ is a network where firms $i$ and $j$ have a link (that is $g_{ij}=1$), and $g^{\prime}$ is the same network $g$, excluding the link between firms $i$ and $j$ (that is $g^{\prime}_{ij}=0$ and $g^{\prime}_{-ij}=g_{-ij}$). Notice that the right-hand side of (\ref{eq:potential_properties}) represents the incentive of firms $i$ and $j$ to form the link; if the sum of their payoffs when they form the link is greater than the sum of their payoffs when they do not have a link, then they will form the link (excluding the stochastic matching shock). The left-hand side of (\ref{eq:potential_properties}) shows that this difference in payoffs can be retrieved using the potential function. This property holds for any pair of firms $i$ and $j$ and for any link $g_{ij}$, $i,j=1,...,n$.\footnote{The formal proof is contained in \cite{Mele2010a} and \cite{MeleZhu2015}.}

This relationship means that the difference in potential functions (the left-hand side of equation (\ref{eq:potential_properties})) can be computed using the profitable deviations of firms $i$ and $j$. Therefore all the equilibrium networks can be found using the potential.

A network is \emph{pairwise stable with transfers} if no two firms want to form a link or delete a link. In our model, all the \emph{pairwise stable  networks (with transfers)}  correspond to the (local) maxima of the potential function (\ref{eq:potential}).\footnote{See \cite{MondererShapley1996}, \cite{Mele2010a}, \cite{JacksonWatts2001} and \cite{Badev2013}.} Intuitively, let's consider a network that maximizes the potential function. Suppose we delete the link between firms $i$ and $j$. This event will decrease the potential because the network was a maximizer of the potential. A decrease in potential means that the sum of payoffs of $i$ and $j$ is higher when the link exists than when we delete their link. Therefore, they will not be willing to delete their partnership. The same will hold if we consider an additional link between two firms in the network.  Adding a link will decrease the potential and therefore implies that the firms involved in this relationship are better off not forming the new link. We can repeat this reasoning for any pair of firms, showing that indeed the network that maximizes the potential function is a pairwise equilibrium network with transfers. This result is important, and it facilitates the computation of the equilibria. Furthermore, at least a pairwise stable equilibrium with transfers is guaranteed by the potential function's existence, as shown in \cite{MondererShapley1996}.

We should note that the potential function is different from a welfare function. Indeed if we assume a classical welfare function, which is the sum of the firms' payoffs, we have
\begin{eqnarray}
W(g,x;\theta) &= & \sum_{i=1}^{n}U_{i}(g,x;\theta) \\
&=& \sum_{i=1}^{n}\sum_{j=1}^{n}g_{ij}u(x_i,x_j;\alpha) + \beta\sum_{i=1}^{n}\sum_{j=1}^{n}\sum_{r\neq i,j}^{n}g_{ij}g_{jr} +
 \gamma\sum_{i=1}^{n}\sum_{j=1}^{n}\sum_{r\neq i,j}^{n}g_{ij}g_{jr}g_{ri}
\label{eq:welfare_fcn}
\end{eqnarray} 

As a consequence, the pairwise stable networks are not necessarily efficient. This is a standard result in the economics of networks literature. The decentralized equilibrium is inefficient because there are externalities in link formation; a firm that forms a direct link pays a cost and receives some benefits, while an indirect link has no cost but brings additional benefits (positive or negative). Therefore, forming a new link creates an externality for other firms that could benefit from a new indirect connection without paying any cost. This externality can be positive or negative, depending on the sign of the coefficients $\beta$ and $\gamma$. \\

%The existence of a potential function is important because it guarantees existence of at least one equilibrium. An additional practical advantage is that one can simulate the network formation process without keeping track of each player profitable deviations: all that information is already incorporated in the potental function, which is a scalar.\footnote{The use of potential games and potential functions is common in computer science and physics. In economics, the class of congestion games is the classical example of potential games.}

Using the potential function characterization we can write the conditional probability of linking as \\
\begin{small}
\begin{eqnarray}
P(g_{ij}=1\vert g_{-ij}, x_i,x_j,\theta) = \frac{\exp\left[\Delta_{ij} \right]}{1+\exp\left[\Delta_{ij} \right]} &=& \notag 
\end{eqnarray}
\begin{eqnarray}
 &=&
\frac{\exp\left[Q(g_{ij}=1,g_{-ij},x;\theta) - Q(g_{ij}=0,g_{-ij},x;\theta) \right]}{1+\exp\left[Q(g_{ij}=1,g_{-ij},x;\theta) - Q(g_{ij}=0,g_{-ij},x;\theta) \right]} \\
&=& \frac{\exp\left[u(x_i,x_j;\alpha) + u(x_j,x_i;\alpha) + \beta\sum_{r\neq i,j}^{n} (g_{jr} + g_{ir}) + \gamma \sum_{r\neq i,j}^{n}(g_{jr}g_{ri} + g_{ir}g_{rj}) \right]}{1 + \exp\left[u(x_i,x_j;\alpha) + u(x_j,x_i;\alpha) + \beta\sum_{r\neq i,j}^{n} (g_{jr} + g_{ir}) + \gamma \sum_{r\neq i,j}^{n}(g_{jr}g_{ri} + g_{ir}g_{rj}) \right]}
\end{eqnarray}
\end{small}
Therefore, the potential fully characterizes the conditional choice probabilities of the firms.\\

The sequence of graphs generated by the network formation game is a Markov chain (\cite{LevinPeresWilmer2008}, \cite{MeynTweedie2009}), converging to a unique stationary equilibrium distribution over networks, which we can characterize in closed-form as
\begin{equation}
\pi(g,x;\theta) = \frac{\exp\left[Q(g,x;\theta)\right]}{c(\theta,x)}
\label{eq:statdist}
\end{equation}
where $Q(g,x;\theta)$ is the potential function in (\ref{eq:potential}) and $c(\theta,x)$ is a normalizing constant that sums over all possible networks with $n$ nodes (we call this set $\mathcal{G}$)
\begin{equation}
c(\theta,x) = \sum_{\omega\in \mathcal{G}}\exp\left[Q(\omega,x;\theta)\right]
\label{eq:normalizing_constant}
\end{equation}
The normalizing constant (\ref{eq:normalizing_constant}) guarantees that equation (\ref{eq:statdist}) is a proper distribution, in other words, that it sums to one when we sum over all possible networks.\footnote{The proof of convergence and the computation of (\ref{eq:statdist}) can be found in \cite{Mele2010a} and \cite{MeleZhu2015}.} 

Equation (\ref{eq:statdist}) is the likelihood of observing a particular network structure in the long run. Notice that this likelihood has peaked at the maxima of the potential function. Therefore, in the long run, we are most likely to observe the networks with high potential. From the discussion above, we know that all the pairwise stable equilibrium networks are maxima of the potential function; as a consequence, the model predicts that, in the long run, we observe equilibrium networks with very high probability.

To perform the estimation, we will assume that the network of firms observed in our data realizes the model's long-run stationary equilibrium. Accordingly, the distribution (\ref{eq:statdist}) is the likelihood of observing a particular network. \\

Finally, our model has a very important property. Let's consider the potential function (\ref{eq:potential}) and its elements. The sum
\begin{equation}
t_{S}(g,x) = \frac{1}{2}\sum_{i=1}^{n}\sum_{j=1}^{n}\sum_{r\neq i,j}^{n}g_{ij}g_{jr} 
\end{equation}
corresponds to the number of 2-stars in network $g$; while the sum
\begin{equation}
t_{T}(g,x)=\frac{2}{3}\sum_{i=1}^{n}\sum_{j=1}^{n}\sum_{r\neq i,j}^{n}g_{ij}g_{jr}g_{ri}
\end{equation}
is the number of triangles in network $g$. Thus the likelihood of the network formation game (\ref{eq:statdist}) corresponds to the likelihood of an ERGM with 2-stars and triangles. We can easily obtain alternative ERGM specifications by changing the structure of the payoffs in Assumption \ref{assumption:utilityscalar}.

\subsection{Extensions of the Network Formation Model}
While this exposition has focused on a dynamic network formation model, characterizing the long-run distribution of networks in equilibrium, we can interpret this model in an alternative fashion. Consider a static network formation game, where each firm \emph{simultaneously} chooses its portfolio of links. In such a game, multiple equilibrium networks are pairwise stable (with transfers). The stationary equilibrium of our model, described in the previous section, corresponds to a refinement of the static equilibrium, called \emph{stochastic best-response dynamics} (\cite{Blume1993}, \cite{JacksonWatts2001}). According to this equilibrium refinement, firm pairs randomly encounter the opportunity to revise one of their link choices, but they make "mistakes" (modeled through random matching shocks $\varepsilon_{ij}$). The iteration of this stochastic best-response procedure generates the long-run distribution of networks (\ref{eq:statdist}). This refinement of the equilibrium concept can also be considered an equilibrium selection device.

The network formation game presented in this paper is a model for a single network observation. We can estimate the parameters $(\alpha, \beta, \gamma)$ by observing only a single network. However, if we have multiple networks or observe the network over time, we can exploit the information in the dynamics to identify richer sets of payoffs.  

We can generalize the payoff functions used in this paper to include more externalities from link formation and accommodate externalities that also depend on each firm's observable characteristics. However, to precisely estimate such payoffs, we need a more extensive network or multiple network observations.\bigskip\\

\section{Estimation details}\label{app:estimation}
We estimated all the models using the package \texttt{Bergm} in R, developed by \cite{CaimoFriel2010}.
All the computations have been performed on a desktop Dell
Precision T7620 with 2 Intel Xeon CPUs E5-2697 v2 with 12 Dual-core processors at 2.7GHZ each and 64GB of RAM.

Fifty thousand parameter simulations yield each table of parameter estimates after discarding 5000 simulations as burn-in. For each parameter proposal, we simulate the model for 10000 iterations and pick the last simulated network to compute the exchange algorithm's acceptance ratio.
We use the snooker algorithm with 10 parallel simulations to improve convergence as implemented in the package (see \cite{CaimoFriel2010}.\\

Codes for estimation are available upon request. 

\subsection{Implementation of the exchange algorithm}
The exchange algorithm shown in the estimation section is computationally intensive and requires some tuning.

The algorithm proceeds as follows. At each iteration $s=1,2,...$ and current parameter $\theta_s$
\begin{enumerate}
	\item Propose a new parameter vector $\theta^{\prime}$
	\begin{equation*}
	\theta^{\prime} \sim q_{\theta}(\cdot\vert \theta_s)
	\end{equation*}
	
	\item Given the proposed parameter $\theta^{\prime}$, simulate a network $g^{\ast}$ as follows. At iteration $r$
	\begin{enumerate}
		\item Propose a new network $g^{\prime}$ 
		\begin{equation*}
			g^{\prime}\sim q_{g}(\cdot\vert g_r)
		\end{equation*} 
		\item Then update the network at iteration $r+1$ 
		\begin{equation*}
			g_{r+1} = \left\lbrace 
			\begin{array}{ll}
			g^{\prime} & \text{with prob. } \alpha_g \\
			g_r & \text{with prob. } 1-\alpha_g
			\end{array}
			\right .
		\end{equation*}
		where $\alpha_g$ is
		\begin{equation*}
		\alpha_g =\min\left\lbrace  1, 
		\frac{\exp\left[Q(g^{\prime},x;\theta^{\prime})\right]}{\exp\left[Q(g_r,x;\theta^{\prime}) \right]} 
		\frac{q_{g}(g_r\vert g^{\prime})}{q_{g}(g^{\prime}\vert g_r)}\right\rbrace
		\end{equation*}
		\item Iterate this process for $r=1,...,R$ and collect the last network $g_R = g^{\ast}$.
	\end{enumerate}
	
	\item Update the parameter at iteration $s+1$ 
	\begin{equation*}
				\theta_{r+1} = \left\lbrace 
			\begin{array}{ll}
			\theta^{\prime} & \text{with prob. } \alpha_{ex} \\
			\theta_r & \text{with prob. } 1-\alpha_{ex}
			\end{array}
			\right .
	\end{equation*} 
	where the probability $\alpha_{ex}$ is given by
	\begin{equation}
	\alpha_{ex} = \min\left\lbrace  1, 
	\frac{\exp\left[Q(g^{\ast},x;\theta^{\prime})\right]}{\exp\left[Q(g_r,x;\theta^{\prime}) \right]} 
	\frac{ \exp\left[Q(g_r,x;\theta_s)\right]}{\exp\left[Q(g^{\ast},x;\theta_s) \right] }  \frac{p(\theta^{\prime}) }{p(\theta_s)}
	\frac{q_{\theta}(\theta_s\vert \theta^{\prime})}{q_{\theta}(\theta^{\prime}\vert \theta_s)}
	\right\rbrace
	\label{eq:exchange_algo_alpha}
	\end{equation}
\end{enumerate}
Notice that the probability (\ref{eq:exchange_algo_alpha}) does not contain any normalizing constant. Neither $\kappa$ nor $c(\theta,x)$ appear in the formulas; therefore, our simulations are feasible. The simulations of parameters converge to the posterior distribution of the structural parameters. All the technical details and the proofs of convergence are in Appendix B of \cite{Mele2010a}. \\

The proposal distribution $q_{\theta}(\cdot\vert \theta_s)$ is a normal centered at the current parameter. To obtain the optimal variance for this proposal distribution, we estimate the model several times and adjust the proposal distribution variance to obtain better acceptance rates. The exchange algorithm has low acceptance rates compared to a standard Metropolis-Hastings algorithm. Our estimates have acceptance rates of $13\%$ and $18\%$, compared with an optimal (asymptotic) rate of $25\%$ for the Metropolis-Hastings.

The proposal distribution for networks $q_{g}(\cdot\vert g_r)$ selects a random pair of nodes and proposes to end their link (if it exists) or form the link (if it does not exist). With a small probability, the proposed network swaps all the entries of the adjacency matrix. This step allows the sampler to reach other modes, extremely useful if the likelihood has multiple modes. For example, this is the case in several models analyzed in \cite{Mele2010a}. \bigskip\\w

\section{Additional estimation results}
This appendix reports alternative estimates and goodness of fit tests for the model. These results were obtained using Maximum Pseudolikelihood estimation and Monte Carlo Maximum Likelihood estimation.
The MPLE results are shown in Table \ref{tab:endog_network_mple}. Notice that this is a frequentist approach and it delivers estimated parameters and standard error. We notice that the estimates are quite different from the posterior means shown in Table \ref{tab:endog_network}, especially the parameters for Number of common partners. 
\begin{table}[ht]
\caption{Model with endogenous network variables, estimated with Maximum Pseudolikelihood (MPLE)}
\centering
\begin{tabular}{l|cc}
\hline\hline

Variable &  Estimate & Std. Error \\

\hline\hline 

Cost           &   -5.078 &  0.039 \\
Number of partners     &0.007  & 0.0001 \\
Common partners       &    0.268 &  0.002  \\
Same firm type     & -0.154 &  0.031    \\
Abs. Difference Capital (log)     & -0.016 &  0.009\\
Abs. Difference Age  &       -0.016 &  0.001  \\
Same state   &          0.510  & 0.037  \\
\hline\hline 

\end{tabular} \\
\label{tab:endog_network_mple}
\end{table}

 The main advantage of the MPLE is that it is quite fast and usually converges relatively fast to the final estimate. This advantage is due to the assumption for the Maximum Pseudolikelihood estimator that the conditional choice probabilities of forming links factorize into the pseudolikelihood: the underlying assumption is that conditional on the rest of the network, the links are independent. This assumption is not satisfied if we believe that the actors (firms) are strategic in their decision to form links. Theoretical results are showing that MPLE estimates are consistent. On the other hand, many practitioners complain of this estimator's poor performance, and the standard errors are imprecise and underestimated.

 To give an idea of this estimator's poor performance, we provide the goodness of fit tests similar to those performed after the Bayesian estimation. We report our tests in Figure \ref{fig:gof_mple}. We simulate 100 networks from the model, using the parameters in Table \ref{tab:endog_network_mple} and compare the observed network to the distribution of the simulated ones. We focus on three network statistics: the degree distribution, the edge-wise shared partners, and the geodesic distance. The solid line represents the observed values in our data, while the dotted lines are the 95$\%$ confidence levels for the simulated networks. The estimated model can replicate the observed network's feature if the solid line is within the dotted lines.

\begin{figure}
\caption{Goodness of fit, model with network effects, Maximum Pseudolikelihood estimates}
\centering
\includegraphics[scale = .75]{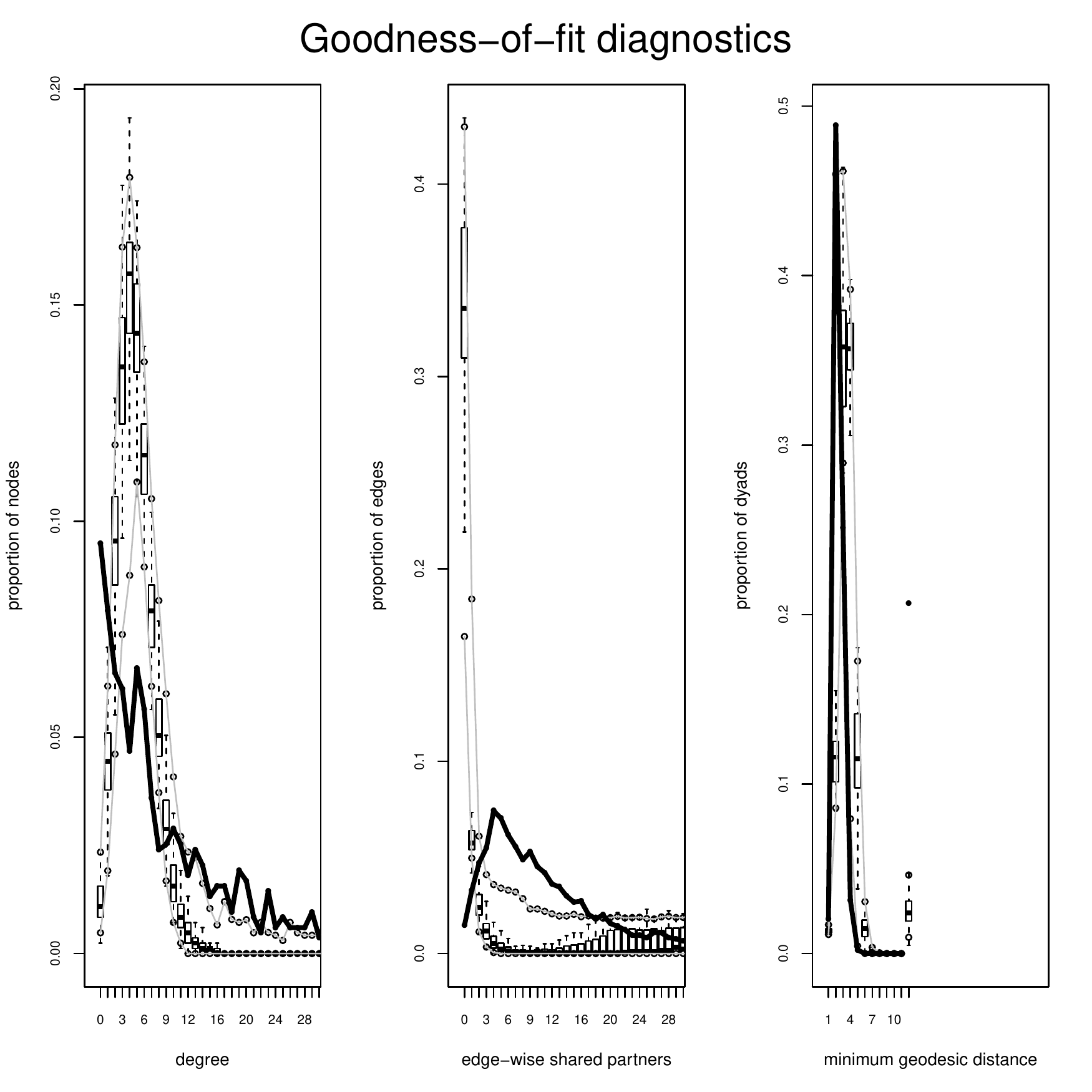}
\flushleft The black solid line represents the observed values of the degree distribution, distribution of edge-wise shared partners and minimu geodesic distance, respectively. The dotted lines represent the 95$\%$ confidence interval for the simulated model. The test is performed by simulating 100 networks from the equilibrium model. The parameters for
the simulations are in Table \ref{tab:endog_network_mple}. Each network is simulated with a MCMC run of 100000 steps.
\label{fig:gof_mple}
\end{figure}
%\subsection{Additional Figures}
%ACF for each parameter\\

The picture shown in Figure \ref{fig:gof_mple} provide evidence that the MPLE is inadequate to fit these data, and performs worse than our Bayesian approach.

%Histograms of the marginal posteriors for the cost of links $\alpha_k$'s are shown in Figure \ref{fig:alphas_posterior}. The marginal posteriors for the covariates are shown in Figure \ref{fig:betas_posterior}. The marginal posterior of the preference for common friends $\gamma_k$'s are in Figure \ref{fig:gammas_posterior}.
%
%\begin{figure}
%\caption{Estimated marginal posterior of cost of links for different communities $(\alpha)$}
%%\includegraphics[scale=.3]{figures/hist_hergm_theta1.pdf}
%%\includegraphics[scale=.3]{figures/hist_hergm_theta2.pdf} \\
%Commmunity 1\hspace{2cm}Community 2\\
%%\includegraphics[scale=.3]{figures/hist_hergm_theta3.pdf}
%%\includegraphics[scale=.3]{figures/hist_hergm_theta4.pdf}\\
%Commmunity 3\hspace{2cm}Between communities
%\label{fig:alphas_posterior}
%\end{figure}
%
%\begin{figure}
%\caption{Estimated marginal posterior of preference parameter for observable characteristics $(\beta)$}
%%\includegraphics[scale=.35]{figures/hist_ergm_theta1.pdf}
%%\includegraphics[scale=.35]{figures/hist_ergm_theta2.pdf}\\
%Same SIC\hspace{5cm}Same country\\
%\label{fig:betas_posterior}
%\end{figure}
%
%
%\begin{figure}
%\caption{Estimated marginal posterior of preference parameter for common friends in different communities $(\gamma)$}
%%\includegraphics[scale=.3]{figures/hist_hergm_theta5.pdf}
%%\includegraphics[scale=.3]{figures/hist_hergm_theta6.pdf} \\
%Commmunity 1\hspace{2cm}Community 2\\
%%\includegraphics[scale=.3]{figures/hist_hergm_theta7.pdf}\\
%Commmunity 3
%\label{fig:gammas_posterior}
%\end{figure}
%
%

  \end{APPENDICES}

% Acknowledgments here
%\ACKNOWLEDGMENT{}

% References here (outcomment the appropriate case)

% CASE 1: BiBTeX used to constantly update the references
%   (while the paper is being written).
%\bibliographystyle{informs2014} % outcomment this and next line in Case 1
%\bibliography{<your bib file(s)>} % if more than one, comma separated

% CASE 2: BiBTeX used to generate mypaper.bbl (to be further fine tuned)
%\input{mypaper.bbl} % outcomment this line in Case 2

%If you don't use BiBTex, you can manually itemize references as shown below.

\bibliographystyle{informs2014}
\bibliography{thesisbib}

%%%%%%%%%%%%%%%%%
\end{document}